\documentclass[superscriptaddress, twocolumn,showpacs,
amssymb,amsmath,nobibnotes,aps,prd,
nofootinbib]{revtex4-2}
\pdfoutput=1
\usepackage{graphicx,subfigure,bm,color,psfrag,hyperref}
\usepackage{amsfonts}
\usepackage{lipsum}
\usepackage{mathtools}
\usepackage{verbatim}
\usepackage[normalem]{ulem}
\usepackage[dvipsnames]{xcolor}
\hypersetup{colorlinks,linkcolor={blue},citecolor={red},urlcolor={magenta}}

\begin{document}

\title{Exploring the Limits: Neutrino Mass and Interacting Dark Sector in the DESI Era}

\author{Weiqiang Yang}
\email{d11102004@163.com}
\affiliation{Department of Physics, Liaoning Normal University, Dalian, 116029, People's Republic of China}

\author{Sibo Zhang}
\email{sbzhang02@163.com}
\affiliation{Department of Physics, Liaoning Normal University, Dalian, 116029, People's Republic of China}

\author{Supriya Pan}
\email{supriya.maths@presiuniv.ac.in}
\affiliation{Department of Mathematics, Presidency University, 86/1 College Street, Kolkata 700073, India}
\affiliation{Institute of Systems Science, Durban University of Technology, Durban 4000, Republic of South Africa}

\author{Luis A. Anchordoqui}
\email{luis.anchordoqui@gmail.com}
\affiliation{Department of Physics and Astronomy,  Lehman College, City University of
  New York, NY 10468, USA}

\affiliation{Department of Physics,
 Graduate Center,  City University of
  New York,  NY 10016, USA}

\affiliation{Department of Astrophysics,
 American Museum of Natural History, NY
 10024, USA}

\author{Eleonora Di Valentino}
\email{e.divalentino@sheffield.ac.uk}
\affiliation{School of Mathematical and Physical Sciences, University of Sheffield, Hounsfield Road, Sheffield S3 7RH, United Kingdom}

\begin{abstract}
\noindent
We constrain three interacting dark matter--dark energy scenarios
using cosmic microwave background measurements, DESI DR2 baryon
acoustic oscillations, and the PantheonPlus, Union3, and DES-Dovekie
Type Ia supernova compilations. We consider interacting vacuum
energy and interacting dark energy with a constant equation of state
in the phantom and quintessence regimes, analyzing each scenario
with either a fixed total neutrino mass or a freely varying
effective neutrino mass.
When the positivity prior is relaxed, all three scenarios favor
negative effective neutrino masses for every dataset combination.
Although some combinations permit small positive masses, none
of the 95\% confidence intervals reaches the minimum mass required
by neutrino oscillation experiments. Allowing the effective neutrino
mass to vary produces significances for a nonzero coupling reaching
approximately $3\sigma$, but the preferred masses remain negative.
These results demonstrate that the interaction and constant
dark energy equations of state considered here fail to resolve
the DESI neutrino-mass tension.
\end{abstract}

\maketitle
\section{Introduction}	

Recent data from the Dark Energy Spectroscopic Instrument (DESI) offer
compelling evidence that dark energy may be dynamical and evolving
over time, rather than acting as a static cosmological
constant~\cite{DESI:2025zgx}. By mapping the spatial distribution of
millions of galaxies and quasars, and combining these measurements with
cosmic microwave background (CMB)~\cite{Planck:2018vyg,AtacamaCosmologyTelescope:2025blo}
and Type Ia supernova (SN Ia)
measurements~\cite{Scolnic:2021amr,Brout:2022vxf,DES:2024jxu}, the
collaboration identified this evolutionary trend at a significance of
approximately $3\sigma$ or higher. Notably, this signal remains resilient
across multiple SN Ia compilations, including the newly recalibrated
DES-Dovekie, Union3.1, and corrected PantheonPlus
datasets~\cite{DES:2025sig,Hoyt:2026fve,Rubin:2026qdt}. Furthermore, the
Dark Energy Survey provided critical validation and support for the
DESI Collaboration's findings~\cite{DES:2026jmi}.

More specifically, the DESI Collaboration examined the
$(w_0,w_a)$ plane by modeling the dark energy equation of state with
a two-parameter, first-order Taylor expansion in the scale factor
about $a=1$, $w(z)=w_0+w_a\,z/(1+z)$, known as the
Chevallier--Polarski--Linder (CPL)
parametrization~\cite{Chevallier:2000qy,Linder:2002et}. Most remarkably,
the constraints point toward an exotic cosmological trajectory wherein
$w(z)$ evolves from the phantom regime ($w<-1$) at higher redshifts,
crosses the phantom barrier ($w=-1$) at intermediate redshifts
($z\sim0.5$), and transitions above $-1$ in the local universe.
These findings have motivated extensive investigations of
dynamical dark energy using different combinations of cosmological
datasets, alternative equation-of-state parametrizations,
nonparametric reconstructions, physical dark energy models,
and modified-gravity scenarios~\cite{DESI:2024mwx,Tada:2024znt,Luongo:2024fww,Cortes:2024lgw,Carloni:2024zpl,Wang:2024dka,Chan-GyungPark:2024mlx,Shlivko:2024llw,Seto:2024cgo,DESI:2024kob,Bhattacharya:2024hep,Gialamas:2024lyw,Najafi:2024qzm,Ye:2024ywg,Giare:2024gpk,Dinda:2024ktd,Jiang:2024xnu,Reboucas:2024smm,RoyChoudhury:2024wri,Giare:2024ocw,Wolf:2025jlc,Giare:2025pzu,Jiang:2025hco,Paliathanasis:2025cuc,Feng:2025mlo,Berti:2025phi,DESI:2025zgx,DESI:2025fii,Paliathanasis:2025dcr,Anchordoqui:2025fgz,Kessler:2025kju,Santos:2025wiv,Yang:2025mws,Wolf:2025jed,Li:2025cxn,Toda:2025dzd,Paliathanasis:2025hjw,RoyChoudhury:2025dhe,Specogna:2025guo,Scherer:2025esj,Cheng:2025lod,Poulin:2025nfb,Cheng:2025hug,vanderWesthuizen:2025iam,Cai:2025mas,Herold:2025hkb,Ozulker:2025ehg,Hogas:2025ahb,Liu:2025vda,Lee:2025pzo,Bhattacharjee:2025xeb,Goldstein:2025epp,Guedezounme:2025wav,Ishak:2025cay,Paul:2025wix,Toda:2025kcq,Luciano:2025ovj,Fazzari:2025lzd,Rezaei:2025vhb,Sabogal:2025qhz,Luciano:2025dhb,Smith:2025icl,Wang:2025djw,Li:2025vuh,Paliathanasis:2025mvy,Zhou:2025ugf,Cheng:2025yue,Zhao:2025rya,Shlivko:2025krk,Garza:2026kys,Ladeira:2026jne,Luciano:2026vhm,Paliathanasis:2026ymi,Zhang:2026cux,Aboubrahim:2026tks,Kang:2026vqj,Turyshev:2026ewm,Li:2026aqz,Gokcen:2026pkq,Najafi:2026kxs,Yang:2026yaq,BeltranJimenez:2026ymd,Sahlu:2026bsa,Shlivko:2026jxa,Paliathanasis:2026vhi,Fernandes:2026zkq,Truong:2026mtr,Paliathanasis:2026acn,Addazi:2026ulq,Toda:2026yum,Paliathanasis:2026dqk,Kessler:2026dbi,Lee:2026yzs,Li:2026asg,Yang:2026kxp,Giare:2026oti,Garcia-Garcia:2026nzy,Liu:2026nds,GuptaChoudhury:2026gsl,Tsilioukas:2026gvy,Carloni:2026ard,Basilakos:2026atu,Jiang:2026srd,Wen:2026xlj,Zhang:2026axl,Zheng:2026tye,Ge:2026ndl,Paliathanasis:2026qka,Sun:2026kig}.

Shortly after the first DESI results came out, it became evident that an
interacting dark matter-dark energy model with an energy transfer rate
proportional to the cosmic expansion rate successfully accommodates
the DESI findings by naturally replicating the apparent dynamical
dark energy behavior~\cite{Giare:2024smz}. Subsequent studies have
further explored comparable interaction rates and their extensions,
examining their observational constraints, theoretical consistency,
and implications for structure
formation~\cite{Pooya:2024wsq,Li:2024qso,Li:2025owk,Silva:2025hxw,Yang:2025uyv,vanderWesthuizen:2025rip,Li:2025muv,Zhang:2025dwu,Pan:2025qwy,Li:2026xaz,Zhai:2026uwr,Yang:2026vkc,Zhang:2026dtd,Sabogal:2026bsw,Li:2026jgg}.
Other studies confirmed that alternative interacting dark sectors
can also explain the unusual dark energy behavior observed by DESI.
Most notably, coupled quintessence fields evolving along
exponential~\cite{Bedroya:2025fwh}
or inverse power-law~\cite{Li:2026xaz} potentials fit the data better
than the $\Lambda$ cold dark matter (CDM) model.\footnote{The
dark-dimension framework~\cite{Montero:2022prj} underlying the coupled quintessence
scenario of~\cite{Bedroya:2025fwh} can also incorporate bulk
Majorana neutrinos~\cite{Anchordoqui:2026ret}.}
This improvement matches or exceeds the statistical significance
achieved by the standard CPL parametrization. In fact, within
these coupled quintessence models, the shape of the potential
has a very small effect on finding a solution that fits the
intriguing DESI results~\cite{Anchordoqui:2026hys}.

A seemingly different, but in fact closely related, subject is the
joint analysis of DESI data combined with Planck and ACT
observations~\cite{Planck:2018vyg,AtacamaCosmologyTelescope:2025blo}
within the standard $\Lambda$CDM framework, which has established
a stringent upper limit on the total neutrino mass of
$\sum m_{\nu}<0.064~{\rm eV}$ at the 95\%
confidence level~\cite{Elbers:2025vlz}
($\sum m_{\nu}<0.055~{\rm eV}$ at the 95\%
confidence level when DESI data are combined with Planck,
ACT, and SPT~\cite{SPT-3G:2025bzu}). These tight cosmological
constraints directly challenge the minimum mass scales determined
by terrestrial neutrino oscillation experiments, which require
approximately $\sum m_{\nu}\geq0.057~{\rm eV}$ for the normal
ordering and $\sum m_{\nu}\geq0.10~{\rm eV}$ for the inverted
ordering~\cite{Esteban:2024eli}.

More acutely, when the physical prior constraint of $\sum m_\nu>0$
is relaxed, the joint dataset exhibits a clear statistical preference
for an unphysical, negative effective neutrino mass
sum~\cite{Craig:2024tky,Green:2024xbb,Loverde:2024nfi}. This anomaly
reflects discrepancies between cosmological datasets propagated
through parameter degeneracies. While CMB-only contours exhibit a
positive correlation between $\sum m_\nu$ and the dimensionless
density parameter for non-relativistic matter $\Omega_m$, this
degeneracy is partially broken by the inclusion of baryon acoustic
oscillation (BAO) distance measurements and SN Ia
data~\cite{Loverde:2024nfi}. Consequently, the preference for a
negative $\sum m_\nu$ signals a mismatch between the late-time
matter fraction inferred from distance measurements and that
inferred from the CMB within the baseline $\Lambda$CDM model.

This conceptual tension underscores potential limitations in the
$\Lambda$CDM framework, offering a strong impetus to explore
alternative cosmological models capable of resolving these
discrepancies~\cite{Craig:2024tky,Green:2024xbb,Elbers:2024sha,Jiang:2024viw,Loverde:2024nfi,Elbers:2025vlz,Graham:2025dqn,Pulido-Hernandez:2026hcs,Yang:2026yaq,Kibris:2026cqq}.
By extending the cosmological model to include a $w_0w_a$CDM
dynamical dark energy parametrization or a matter-to-dark energy
conversion mechanism~\cite{Graham:2025dqn}, the severe neutrino
mass bounds imposed by DESI are significantly relaxed. This allows
for positive neutrino masses that successfully reconcile
cosmological datasets with the lower bounds from terrestrial
oscillation experiments~\cite{Elbers:2025vlz,Yang:2026yaq}.
For example, adopting the CPL 
parametrization broadens the upper bound substantially, yielding
$\sum m_\nu<0.163~{\rm eV}$ at the 95\% confidence level for
the DESI DR2 BAO and CMB combination considered by
Elbers et al.~\cite{Elbers:2025vlz}.\footnote{It is worth
mentioning that another possibility is to exclude the low-$\ell$
CMB polarization data, allowing a higher value of the reionization
optical depth $\tau$~\cite{Giare:2023ejv,Jhaveri:2025neg}.}

Crucially, the geometric component of the BAO-CMB-SN Ia discordance
with positive neutrino masses is physically driven by the matter-era
distance interval spanning from photon-baryon decoupling to the
intermediate redshifts probed by DESI~\cite{Weiner:2026sfm}.
This insight redefines how preferences for the $(w_0,w_a)$
parametrization must be interpreted. Specifically, dark energy dynamics
confined to low redshifts are fundamentally insufficient to resolve
this geometric tension. Instead, within dynamical dark energy
explanations, reconciliation of this matter-era distance excess
requires the dark energy density to be substantially suppressed
relative to a cosmological constant at high redshifts, particularly
at $z>2.33$~\cite{Weiner:2026sfm}. This high-redshift density
suppression represents the physical mechanism by which the
$(w_0,w_a)$ framework successfully relaxes the neutrino mass
bounds~\cite{Costa:2025kwt,Weiner:2026sfm}. Notably, this cosmological
tension is not driven by the choice of a single supernova sample;
neutrino mass constraints from DESI DR2 demonstrate remarkable
stability across different SN Ia compilations, including DES-Dovekie,
and this behavior persists even when supernova data are excluded
entirely from the joint DESI and CMB
analysis~\cite{Yang:2026yaq}.

Given the evolution predicted by various parametrizations of the
equation of state, it is worthwhile to examine whether more realistic
models with effective phantom crossing can concurrently address
the neutrino mass problem. A very recent study took a step in this
direction by analyzing a quintessence field with an exponential
potential and an exponential coupling to dark matter, demonstrating
that its phantom-crossing branch yields effective neutrino mass
constraints consistent with neutrino oscillation data within
$2\sigma$~\cite{Wang:2026gvg}. In this study, we evaluate
three alternative scenarios featuring non-gravitational coupling
within the dark sector. First, we examine a basic scenario in which
vacuum energy with an equation of state $w_x=-1$ interacts with
pressureless dark
matter~\cite{Kumar:2016zpg,Yang:2019bpr,Pan:2022qrr}.
We then consider interacting dark matter-dark energy models with
a constant dark energy equation of state $w_x=w_0\neq-1$,
separately examining the phantom and quintessence
regimes~\cite{Yang:2018euj,Pan:2019gop,Yang:2020uga,DiValentino:2020kpf,Yang:2021hxg,Anchordoqui:2021gji,Giare:2024ytc}.\footnote{A
minimal extension introduces a one-parameter redshift-dependent
equation of state for dark energy~\cite{Yang:2018uae}. We exclude
this approach from our analysis.}
Ultimately, we demonstrate that none of these alternative frameworks
accommodates a positive neutrino mass sum consistent with the lower
bounds established by oscillation experiments at the 95\% confidence
level for the dataset combinations analyzed.

A point worth noting at this juncture is that our analysis
differs from those carried out
in~\cite{DESI:2025ffm,Du:2025xes,Li:2026ldf}, which restrict
the neutrino mass sum to non-negative values, whereas we allow
the effective neutrino mass sum to take negative values and
test consistency with the lower bound from neutrino oscillation
experiments.

The remainder of the paper is structured into four sections. First,
Sec.~\ref{sec:2} summarizes the cosmological background and linear
perturbations of the interacting dark energy models analyzed herein.
Next, Sec.~\ref{sec:3} outlines the datasets and methodology used.
Section~\ref{sec:4} is dedicated to parameter constraints and their
implications. Lastly, Sec.~\ref{sec:5} summarizes our conclusions.

\section{Cosmological Background and Linear Perturbations}
\label{sec:2}

This section outlines the cosmological background and linear
perturbations governing the interaction between cold dark matter ($c$)
and dark energy ($x$) within the framework of general relativity and a
flat Friedmann-Lema\^{\i}tre-Robertson-Walker universe.

We work in the synchronous gauge, adopting the line element
\begin{equation}
ds^2 = a^2(\eta)\left[-d\eta^2+(\delta_{ij}+h_{ij})dx^i dx^j\right]\,,
\end{equation}
where $a(\eta)$ is the scale factor, $\eta$ is conformal time,
$\delta_{ij}$ is the Kronecker delta, and $h_{ij}$ is the perturbation
of the spatial metric.

The energy-momentum conservation equations for $c$ and $x$ include an
interaction term $Q$, which facilitates energy transfer between the
two fluids~\cite{Valiviita:2008iv,Gavela:2009cy,Gavela:2010tm,Yang:2021hxg}:

\begin{equation}
\nabla_{\nu}T_c^{\mu\nu}=-\frac{Qu_c^{\mu}}{a(\eta)},
\end{equation}
\begin{equation}
\nabla_{\nu}T_x^{\mu\nu}=\frac{Qu_c^{\mu}}{a(\eta)}\,,
\end{equation}
where $T_c^{\mu\nu}$ and $T_x^{\mu\nu}$ are the energy-momentum
tensors for $c$ and $x$, respectively, and
$u_c^{\mu}=a^{-1}(\eta)(1,v_c^i)$ is the cold dark matter
four-velocity to linear order, with $v_c^i$ its peculiar velocity.
The transfer four-vector is aligned with the cold dark matter
four-velocity, so there is no momentum transfer in the cold dark
matter rest frame.
The sign of $Q$ dictates the direction of energy transfer:
$Q>0$ transfers energy from $c$ to $x$, while $Q<0$
transfers energy from $x$ to $c$.

To specify the coupling $Q$, we adopt a widely studied
phenomenological interaction rate,
\begin{equation}
Q=3\xi\mathcal{H}\rho_x\,,
\end{equation}
where $\mathcal{H}$ is the conformal Hubble parameter and $\xi$
is a dimensionless, time-independent coupling constant.

We now turn to the linear perturbations. First, we consider the
interacting vacuum scenario ({\bf IVS}) with $w_x=-1$.
Assuming vanishing dark energy density perturbations in the dark
matter comoving frame ($\delta_x=0$), the evolution equations for
the cold dark matter density contrast $\delta_c$ and velocity
divergence $\theta_c$ are, respectively,
\begin{equation}
\delta_c^{\prime}
=-\left(\theta_c+\frac{h^{\prime}}{2}\right)
+3\mathcal{H}\xi\frac{\rho_x}{\rho_c}\delta_c,
\end{equation}
and
\begin{equation}
\theta_c^{\prime}=-\mathcal{H}\theta_c\,,
\label{thetacprime}
\end{equation}
where $h$ is the trace of $h_{ij}$ and primes denote derivatives
with respect to conformal time.

Second, we consider the interacting dark energy ({\bf IDE}) model with
$w_x=w_0\neq-1$. Depending on the nature of the dark energy equation of state, we have considered two distinct scenarios, namely {\bf IDEphan} in which $w_0$ lies in the phantom regime (i.e. $w_0 <-1$), and {\bf IDEquin} in which $w_0$ is quintessential ($w_0 > -1$). 
In this case, the evolution equations for the
dark energy density contrast and velocity divergence, and the cold
dark matter density contrast, can be expressed as
\begin{eqnarray}
\delta_x^{\prime}
&=&-(1+w_x)\left(\theta_x+\frac{h^{\prime}}{2}\right)
-3\mathcal{H}w_x^{\prime}\frac{\theta_x}{k^2}
-3\mathcal{H}(c_{s,x}^2-w_x)\nonumber\\
&\times&
\left[\delta_x+3\mathcal{H}(1+w_x)\frac{\theta_x}{k^2}\right]
+9\mathcal{H}^2\xi(c_{s,x}^2-w_x)\frac{\theta_x}{k^2},
\end{eqnarray}
\begin{eqnarray}
\theta_x^{\prime}
&=&-\mathcal{H}(1-3c_{s,x}^2)\theta_x
+\frac{c_{s,x}^2}{1+w_x}k^2\delta_x
+3\mathcal{H}\xi\nonumber\\
&\times&
\left[\frac{\theta_c-(1+c_{s,x}^2)\theta_x}{1+w_x}\right],
\end{eqnarray}
and
\begin{equation}
\delta_c^{\prime}
=-\left(\theta_c+\frac{h^{\prime}}{2}\right)
+3\mathcal{H}\xi\frac{\rho_x}{\rho_c}(\delta_c-\delta_x),
\end{equation}
with $\theta_c^{\prime}$ given by Eq.~(\ref{thetacprime}).
Here, $k$ is the comoving wave number and $c_{s,x}^2$ is the square of the sound speed in the 
dark energy rest-frame. We set $c_{s,x}^2 =1$ for two reasons: for $c_{s,x}^2 \sim 1$, dark energy does not cluster at the sub-horizon scales and this value is well motivated from scalar field theory.

\section{Methodology and datasets}
\label{sec:3}  	

This section outlines the observational datasets and numerical methods
used to constrain the parameter space of the {\bf IVS} and IDE models.
We use $\Lambda$CDM as the baseline model, which contains six
cosmological parameters:
\begin{equation}
\bm{\theta}=\Bigl\{\Omega_bh^2,\,\Omega_ch^2,\,
100\theta_{\rm MC},\,\tau,\,n_s,\,\ln(10^{10}A_s)\Bigr\}\,,
\end{equation}
where $\Omega_bh^2$ is the baryon density parameter multiplied by
the square of the reduced Hubble constant $h$, $\Omega_ch^2$ is the
CDM density parameter also multiplied by $h^2$, $\theta_{\rm MC}$
is an approximation to the angular size of the sound horizon at
recombination, $\tau$ is the Thomson scattering optical depth due to
reionization, $n_s$ is the scalar spectral index, and $A_s$ is the
amplitude of the primordial power spectrum of adiabatic scalar
perturbations.\footnote{The actual angular sound horizon scale $\theta_s$
is calculated numerically, whereas parameter sampling uses
$\theta_{\rm MC}$ as an approximation based on a fitting formula
for the redshift of recombination.}

We first compare $\Lambda$CDM with the {\bf IVS} and
{\bf IVS} + $\sum m_{\nu,{\rm eff}}$ models. For the {\bf IVS}, the total
neutrino mass is fixed at $\sum m_\nu=0.06~{\rm eV}$.
For the {\bf IVS} + $\sum m_{\nu,{\rm eff}}$ analysis, we introduce an
effective neutrino mass sum, $\sum m_{\nu,{\rm eff}}$, by relaxing
the prior to allow negative values. Thus, the {\bf IVS} has seven free
parameters,
\begin{equation}
\bm{\theta}^{\rm IVS}
=\Bigl\{\bm{\theta},\,\xi\Bigr\}\,,
\end{equation}
and {\bf IVS} + $\sum m_{\nu,{\rm eff}}$ has eight free parameters,
\begin{equation}
\bm{\theta}_{\nu}^{\rm IVS}
=\Bigl\{\bm{\theta},\,\xi,\,\sum m_{\nu,{\rm eff}}\Bigr\}\,.
\end{equation}
We then repeat the procedure for the IDE models. The IDE models
have eight free parameters,
\begin{equation}
\bm{\theta}^{\rm IDE}
=\Bigl\{\bm{\theta},\,\xi,\,w_0\Bigr\}\,,
\end{equation}
and the IDE + $\sum m_{\nu,{\rm eff}}$ models have nine free
parameters,
\begin{equation}
\bm{\theta}_{\nu}^{\rm IDE}
=\Bigl\{\bm{\theta},\,\xi,\,w_0,\,
\sum m_{\nu,{\rm eff}}\Bigr\}\,.
\end{equation}
In all analyses, the effective number of relativistic neutrino
species at high redshift remains fixed at $N_{\rm eff}=3.04$.

We performed Markov chain Monte Carlo (MCMC) analyses using
{\tt CLASS}~\cite{Blas:2011rf,Lesgourgues:2011rh} and
{\tt COBAYA}~\cite{Torrado:2020dgo} to constrain the parameter
space using the datasets listed below. To assess convergence,
we applied the Gelman-Rubin criterion, requiring $R-1<0.02$
for all runs included in our final results. The uniform priors
for each cosmological parameter are listed in
Table~\ref{tab:priors}.

To assess the statistical viability of the {\bf IVS} and IDE models,
with and without a free effective neutrino mass sum, we employ
both minimum chi-squared ($\chi^2_{\rm min}$) and Bayesian
evidence comparisons against $\Lambda$CDM. In each comparison,
the reference $\Lambda$CDM model adopts the same neutrino mass
treatment as the interacting model: either fixed at
$\sum m_\nu=0.06~{\rm eV}$ or with
$\sum m_{\nu,\rm eff}$ varied freely.
The improvement in the best fit is evaluated via
\begin{equation}
\Delta\chi^2_{\rm min}
=\chi^2_{\rm min}(\text{Model})
-\chi^2_{\rm min}(\text{Reference})\,,
\end{equation}
where a negative value indicates a better fit for the {\bf IVS}/{\bf IDE}
model than for the corresponding baseline.

For the Bayesian analysis, the model evidence is computed via
\begin{equation}
\mathcal{B}_i = \int
\mathcal{L}(D\vert{}\Theta,\mathcal{M}_i)\,
\pi(\Theta\vert{}\mathcal{M}_i)\,d\Theta\,,
\end{equation}
where $\mathcal{L}$ is the likelihood of the data $D$ given the
parameters $\Theta$, and $\pi$ is the prior distribution. We quantify
relative model performance using the log Bayes factor,
$\ln \mathcal{B}_{ij} = \ln \mathcal{B}_i - \ln \mathcal{B}_j$,
where $i$ and $j$ represent the candidate and reference models,
respectively. Positive values indicate support for the proposed
framework over $\Lambda$CDM, while negative values indicate support
for $\Lambda$CDM. We estimate the Bayesian evidence from the MCMC
posterior samples using the learned harmonic mean estimator
implemented in \texttt{harmonic}~\cite{harmonic,Polanska:2024arc}.
We interpret the strength of the log Bayes factor according to
the revised Jeffreys' scale~\cite{Kass:1995loi,Trotta:2008qt}:
inconclusive ($0 \leq |\ln \mathcal{B}_{ij}| < 1$),
weak ($1 \leq |\ln \mathcal{B}_{ij}| < 2.5$),
moderate ($2.5 \leq |\ln \mathcal{B}_{ij}| < 5$),
strong ($5 \leq |\ln \mathcal{B}_{ij}| < 10$), and
very strong ($|\ln \mathcal{B}_{ij}| \geq 10$).

\begin{table}
    \centering
    \renewcommand{\arraystretch}{1.4}
    \begin{tabular}{|c|c|c|c|}
        \hline\hline
                \textbf{Parameter} &
        \textbf{Prior}$_{\mathrm{IVS}}$ &
        \textbf{Prior}$_{\mathrm{IDEphan}}$ &
        \textbf{Prior}$_{\mathrm{IDEquin}}$ \\
        \hline\hline
        $\Omega_b h^2$ &
        $[0.005,0.1]$ & $[0.005,0.1]$ & $[0.005,0.1]$ \\
        $\Omega_c h^2$ &
        $[0.001,0.99]$ & $[0.001,0.99]$ & $[0.001,0.99]$ \\
        $\tau$ &
        $[0.01,0.8]$ & $[0.01,0.8]$ & $[0.01,0.8]$ \\
        $n_s$ &
        $[0.8,1.2]$ & $[0.8,1.2]$ & $[0.8,1.2]$ \\
        $\ln(10^{10}A_s)$ &
        $[1.61,3.91]$ & $[1.61,3.91]$ & $[1.61,3.91]$ \\
        $100\theta_{\rm MC}$ &
        $[0.5,10]$ & $[0.5,10]$ & $[0.5,10]$ \\
        $\sum m_{\nu,\mathrm{eff}}$ [eV] &
        $[-1.5,1.5]$ & $[-1.5,1.5]$ & $[-1.5,1.5]$ \\
        $w_0$ &
        $-$ & $[-3,-1]$ & $[-1,0]$ \\
        $\xi$ &
        $[-1,1]$ & $[-1,0]$ & $[0,1]$ \\
        \hline\hline
    \end{tabular}
    \caption{Uniform priors used for the cosmological parameters
    varied independently during the statistical analysis.
    The neutrino mass prior applies only when the effective
    neutrino mass is varied.}
    \label{tab:priors}
\end{table}

In all data analyses, we provide results for additional
cosmological parameters, including: $\sigma_8$, which measures the
standard deviation of the linear matter density contrast at $z=0$
smoothed within spheres of comoving radius $8h^{-1}$~Mpc;
the combined parameter $S_8 = \sigma_8(\Omega_m/0.3)^{0.5}$,
which combines $\sigma_8$ with the total matter density parameter
$\Omega_m$ to capture their degeneracy in weak-lensing measurements;
the Hubble constant
$H_0 = 100h~{\rm km\,s^{-1}\,Mpc^{-1}}$;
and the comoving sound horizon at the drag epoch,
$r_{\rm drag}$, a critical standard ruler defined by the comoving
distance sound waves travel through the photon-baryon fluid before
baryons are released from photon drag.

We base our work on the following observational datasets:

\begin{enumerate}
    \item The Planck NPIPE (PR4) \texttt{CamSpec} likelihood
    for the high-$\ell$ TT, TE, and EE power
    spectra~\cite{Rosenberg:2022sdy}, the \texttt{Commander}
    likelihood for the low-$\ell$ TT spectrum, and the
    \texttt{SimALL} likelihood for the low-$\ell$ EE
    spectrum~\cite{Planck:2018vyg}. Additionally, we include the Planck NPIPE (PR4) lensing and ISW-lensing likelihoods~\cite{Carron:2022eyg,Carron:2022eum}, built from Planck NPIPE data. We refer to this combination as {\bf CMB}.

    \item BAO scale measurements from the second data release
    (DR2) of the DESI survey~\cite{DESI:2025zgx}, providing
    cosmological distance constraints. Hereafter, we denote
    this BAO dataset as \textbf{DESI}.

    \item SN Ia datasets from three sources:
    (i) the \textbf{PantheonPlus}
compilation~\cite{Scolnic:2021amr,Brout:2022vxf};
    (ii) the expanded \textbf{Union3} compilation containing
    2,087 supernovae~\cite{Rubin:2023ovl}; and
    (iii) a new reanalysis of the five-year supernova survey
    data from the \textbf{Dark Energy Survey}
    (\textbf{DES-Dovekie})~\cite{DES:2024jxu,DES:2025sig}.
\end{enumerate}

\begingroup
\begin{center}
	\begin{table*}
		\scalebox{0.85}{				
			\begin{tabular}{ccccccccc}
				\hline
				Parameters & CMB+DESI & CMB+DESI+PantheonPlus & CMB+DESI+Union3 & CMB+DESI+DES-Dovekie\\ \hline

				$\Omega_\mathrm{b} h^2$ & $0.02228_{-0.00013-0.00025}^{+0.00012+0.00025}$ & $0.02232_{-0.00012-0.00025}^{+0.00013+0.00024}$ & $0.02232_{-0.00012-0.00028}^{+0.00014+0.00025}$ & $0.02232_{-0.00012-0.00024}^{+0.00012+0.00024}$ \\

				$\Omega_\mathrm{c} h^2$ & $0.101_{-0.011-0.024}^{+0.013+0.021}$ & $0.1209_{-0.0063-0.014}^{+0.0071+0.013}$ & $0.1203_{-0.0077-0.019}^{+0.0091+0.016}$ & $0.1221_{-0.0057-0.012}^{+0.0066+0.012}$ \\
				
				$100\theta_\mathrm{MC}$ & $1.04191_{-0.00024-0.00046}^{+0.00023+0.00046}$ & $1.04194_{-0.00023-0.00044}^{+0.00023+0.00043}$ & $1.04193_{-0.00023-0.00046}^{+0.00024+0.00044}$ & $1.04193_{-0.00023-0.00045}^{+0.00022+0.00044}$ \\
				
				$\tau$ & $0.0569_{-0.0072-0.014}^{+0.0065+0.015}$ & $0.0594_{-0.0077-0.014}^{+0.0066+0.016}$ & $0.0600_{-0.0078-0.013}^{+0.0066+0.015}$ & $0.0596_{-0.0079-0.014}^{+0.0064+0.016}$ \\
				
				$n_\mathrm{s}$ & $0.9668_{-0.0036-0.0069}^{+0.0035+0.0069}$ & $0.9683_{-0.0035-0.0068}^{+0.0035+0.0065}$ & $0.9684_{-0.0036-0.0072}^{+0.0036+0.0073}$ & $0.9683_{-0.0036-0.0068}^{+0.0035+0.0069}$ \\
				
				$\ln(10^{10} A_\mathrm{s})$ & $3.044_{-0.014-0.026}^{+0.013+0.028}$ & $3.048_{-0.015-0.028}^{+0.013+0.031}$ & $3.048_{-0.016-0.028}^{+0.014+0.030}$ & $3.048_{-0.015-0.026}^{+0.013+0.030}$ \\
				
				$\xi$ & $0.049_{-0.037-0.064}^{+0.034+0.067}$ & $-0.011_{-0.023-0.044}^{+0.022+0.045}$ & $-0.009_{-0.028-0.056}^{+0.028+0.059}$ & $-0.015_{-0.021-0.040}^{+0.021+0.042}$ \\
				
				$\Omega_\mathrm{m}$ & $0.258_{-0.030-0.061}^{+0.033+0.056}$ & $0.313_{-0.019-0.039}^{+0.019+0.038}$ & $0.311_{-0.024-0.053}^{+0.025+0.050}$ & $0.316_{-0.017-0.036}^{+0.019+0.034}$ \\
				
				$\sigma_8$ & $0.921_{-0.11-0.16}^{+0.064+0.19}$ & $0.789_{-0.044-0.075}^{+0.033+0.080}$ & $0.794_{-0.056-0.094}^{+0.040+0.11}$ & $0.783_{-0.039-0.066}^{+0.030+0.071}$ \\
				
				$H_0~[{\rm Km/s/Mpc}]$ & $69.40_{-0.92-1.6}^{+0.84+1.7}$ & $67.88_{-0.56-1.1}^{+0.56+1.1}$ & $67.92_{-0.70-1.5}^{+0.72+1.5}$ & $67.77_{-0.52-0.98}^{+0.52+1.0}$ \\
				
				$S_8$ & $0.848_{-0.038-0.057}^{+0.024+0.067}$ & $0.804_{-0.018-0.032}^{+0.015+0.033}$ & $0.805_{-0.021-0.037}^{+0.017+0.041}$ & $0.802_{-0.016-0.028}^{+0.014+0.031}$ \\
				
				$r_{\rm{drag}}~[{\rm Mpc}]$ & $147.66_{-0.20-0.41}^{+0.20+0.41}$ & $147.80_{-0.20-0.39}^{+0.20+0.40}$ & $147.80_{-0.22-0.39}^{+0.20+0.42}$ & $147.80_{-0.20-0.37}^{+0.19+0.40}$ \\
				
				\hline
				
				$\xi~(\sigma)$ & $1.4$ & $0.50$ & $0.33$ & $0.70$ \\
				
				$\Delta\chi^2_{\rm min}$ & $-1.9$ & $-0.3$ & $-0.6$ & $-1.2$ \\
				
				$\rm{ln}\mathcal{B}_{ij}$ & $-0.5$ & $-1.6$ & $-1.3$ & $-1.2$ \\
				
				\hline 
								
				$\Omega_\mathrm{b} h^2$ & $0.02227_{-0.00014-0.00027}^{+0.00013+0.00025}$ & $0.02228_{-0.00013-0.00025}^{+0.00013+0.00026}$ & $0.02229_{-0.00013-0.00024}^{+0.00012+0.00026}$ & $0.02228_{-0.00013-0.00025}^{+0.00013+0.00025}$ \\
				
				$\Omega_\mathrm{c} h^2$ & $0.128_{-0.015-0.028}^{+0.015+0.028}$ & $0.1377_{-0.0096-0.019}^{+0.0099+0.019}$ & $0.143_{-0.010-0.022}^{+0.012+0.021}$ & $0.1397_{-0.010-0.017}^{+0.0087+0.019}$ \\
				
				$100\theta_\mathrm{MC}$ & $1.04178_{-0.00023-0.00044}^{+0.00023+0.00044}$ & $1.04178_{-0.00023-0.00046}^{+0.00024+0.00046}$ & $1.04177_{-0.00024-0.00046}^{+0.00024+0.00046}$ & $1.04178_{-0.00023-0.00048}^{+0.00023+0.00048}$ \\
				
				$\tau$ & $0.0492_{-0.0076-0.016}^{+0.0078+0.016}$ & $0.0483_{-0.0074-0.017}^{+0.0084+0.016}$ & $0.0479_{-0.0073-0.017}^{+0.0085+0.016}$ & $0.0481_{-0.0072-0.016}^{+0.0081+0.015}$ \\
				
				$n_\mathrm{s}$ & $0.9657_{-0.0037-0.0067}^{+0.0034+0.0073}$ & $0.9660_{-0.0034-0.0075}^{+0.0038+0.0069}$ & $0.9664_{-0.0036-0.0072}^{+0.0036+0.0072}$ & $0.9659_{-0.0036-0.0073}^{+0.0036+0.0069}$ \\
				
				$\ln(10^{10} A_\mathrm{s})$ & $3.028_{-0.015-0.032}^{+0.017+0.033}$ & $3.025_{-0.015-0.034}^{+0.017+0.031}$ & $3.024_{-0.015-0.033}^{+0.017+0.032}$ & $3.025_{-0.015-0.031}^{+0.015+0.031}$ \\
				
				$\sum m_\mathrm{\nu,eff}~{\rm [eV]}$ & $-0.125_{-0.078-0.16}^{+0.083+0.16}$ & $-0.160_{-0.071-0.16}^{+0.082+0.15}$ & $-0.182_{-0.081-0.16}^{+0.078+0.16}$ & $-0.172_{-0.073-0.16}^{+0.080+0.15}$ \\
				
				$\xi$ & $-0.030_{-0.044-0.097}^{+0.050+0.086}$ & $-0.062_{-0.030-0.070}^{+0.036+0.063}$ & $-0.081_{-0.041-0.077}^{+0.040+0.081}$ & $-0.068_{-0.028-0.067}^{+0.037+0.059}$ \\
				
				$\Omega_\mathrm{m}$ & $0.318_{-0.037-0.070}^{+0.038+0.070}$ & $0.343_{-0.024-0.047}^{+0.024+0.046}$ & $0.358_{-0.024-0.056}^{+0.032+0.052}$ & $0.348_{-0.023-0.042}^{+0.023+0.043}$ \\
				
				$\sigma_8$ & $0.797_{-0.080-0.12}^{+0.050+0.14}$ & $0.749_{-0.042-0.072}^{+0.032+0.076}$ & $0.729_{-0.049-0.076}^{+0.031+0.088}$ & $0.742_{-0.038-0.064}^{+0.033+0.072}$ \\
				
				$H_0~[{\rm Km/s/Mpc}]$ & $68.29_{-0.98-1.9}^{+0.99+1.9}$ & $67.63_{-0.66-1.3}^{+0.66+1.3}$ & $67.21_{-0.84-1.5}^{+0.83+1.6}$ & $67.51_{-0.61-1.3}^{+0.70+1.2}$ \\
				
				$S_8$ & $0.815_{-0.027-0.043}^{+0.018+0.050}$ & $0.800_{-0.017-0.029}^{+0.014+0.031}$ & $0.793_{-0.018-0.031}^{+0.014+0.033}$ & $0.798_{-0.015-0.029}^{+0.014+0.030}$ \\
				
				$r_{\rm{drag}}~[{\rm Mpc}]$ & $147.58_{-0.21-0.42}^{+0.21+0.43}$ & $147.62_{-0.21-0.42}^{+0.21+0.42}$ & $147.67_{-0.22-0.44}^{+0.22+0.45}$ & $147.63_{-0.22-0.42}^{+0.21+0.42}$ \\
				
				\hline
				
				$\xi~(\sigma)$ & $0.62$ & $1.9$ & $2.0$ & $2.5$ \\ 
				
				$\Delta\chi^2_{\rm min}$ & $-0.2$ & $-5.4$ & $-5.9$ & $-5.4$ \\
				
				$\rm{ln}\mathcal{B}_{ij}$ & \boldmath{$+0.1$} & \boldmath{$+1.6$} & \boldmath{$+2.4$} & \boldmath{$+2.2$} \\
				
				\hline
		\end{tabular}}                                                                
		\caption{68\% and 95\% CL constraints on the \textbf{IVS}
scenario are presented for the CMB+DESI,
CMB+DESI+PantheonPlus, CMB+DESI+Union3, and
CMB+DESI+DES-Dovekie datasets. The upper half of the table
corresponds to the scenario with the neutrino mass fixed,
while the lower half corresponds to the scenario including
the effective neutrino mass as a free parameter.
We also report the statistical significance of a nonzero
coupling ($\xi$), obtained from its one-dimensional Gaussian
posterior using Wilks' theorem~\cite{Wilks:1938dza}.
This quantifies the statistical evidence for a nonzero value
of the coupling parameter, i.e., the deviation of $\xi$ from zero.}
		\label{tab:IVS}                   
	\end{table*}                                     
\end{center}
\endgroup
\begin{figure*}
	\includegraphics[width=0.8\textwidth]{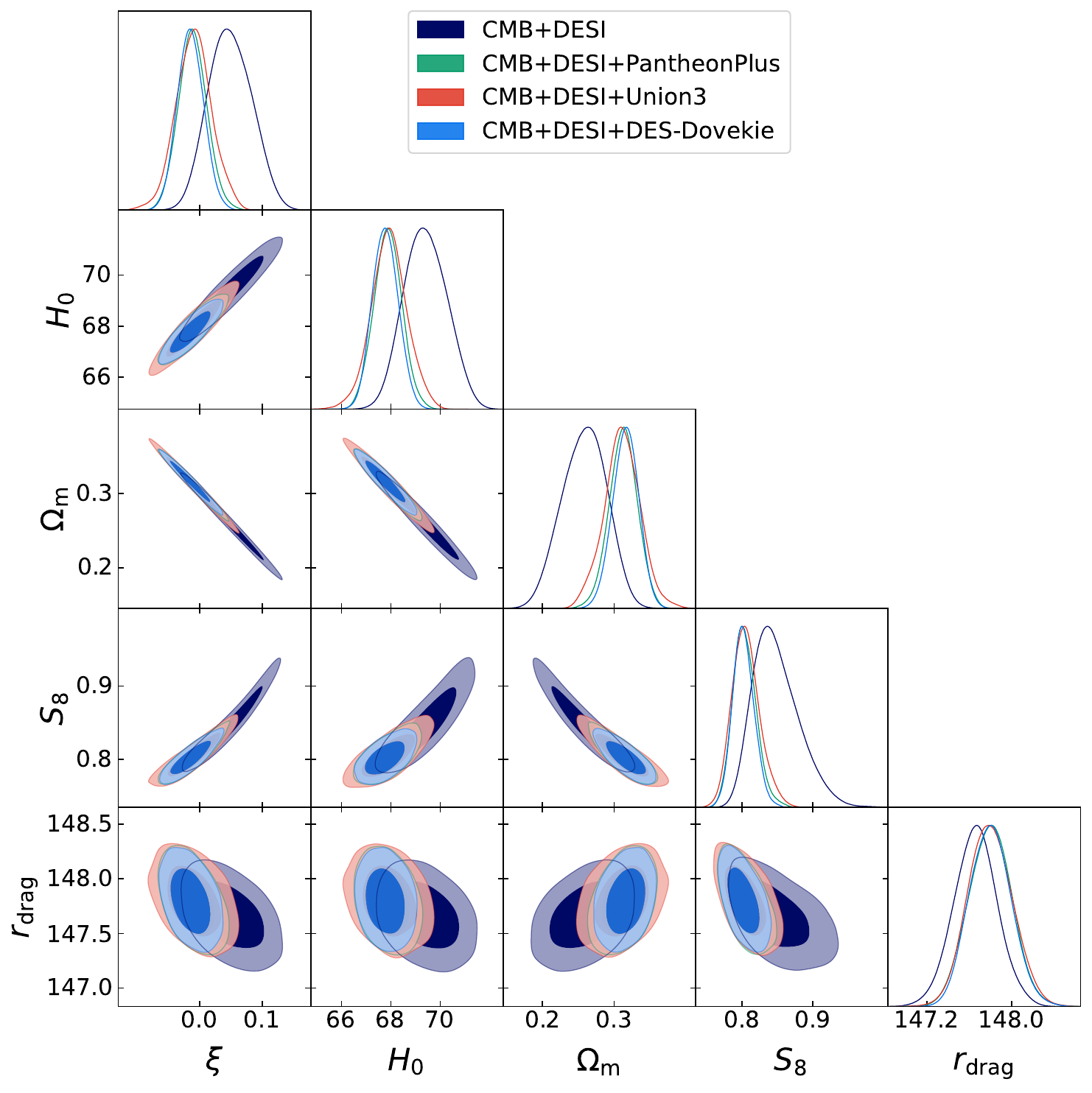}
	\caption{One-dimensional posterior distributions and
two-dimensional joint contours for the most relevant parameters
of the model with fixed neutrino mass, using different
combinations of cosmological measurements.}
	\label{fig:IVS}
\end{figure*}
\begin{figure*}
	\includegraphics[width=0.8\textwidth]{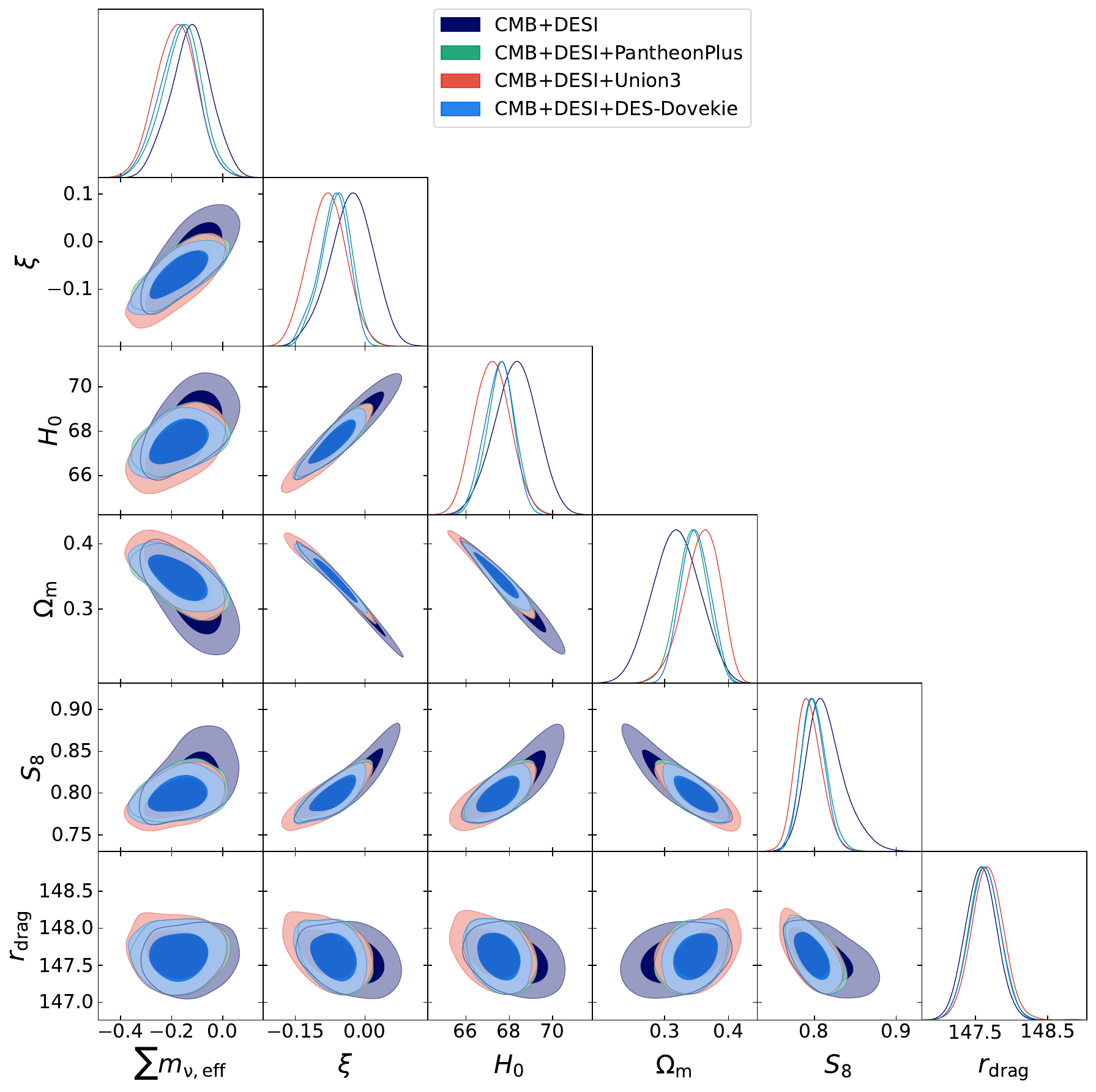}
	\caption{One-dimensional posterior distributions and
two-dimensional joint contours for the most relevant parameters
of the model, using different combinations of cosmological
measurements.}
	\label{fig:IVS_NegMnu} 
\end{figure*}
\begingroup
\begin{center}
\begin{table*}
\scalebox{0.85}{
\begin{tabular}{ccccccccc}
\hline
Parameters & CMB+DESI & CMB+DESI+PantheonPlus & CMB+DESI+Union3 & CMB+DESI+DES-Dovekie\\
\hline

$\Omega_\mathrm{b} h^2$ &
$0.02227_{-0.00014-0.00026}^{+0.00013+0.00026}$ &
$0.02231_{-0.00013-0.00025}^{+0.00013+0.00025}$ &
$0.02231_{-0.00013-0.00025}^{+0.00013+0.00025}$ &
$0.02232_{-0.00013-0.00025}^{+0.00013+0.00025}$ \\

$\Omega_\mathrm{c} h^2$ &
$0.1369_{-0.017-0.019}^{+0.0079+0.024}$ &
$0.1381_{-0.016-0.020}^{+0.0096+0.023}$ &
$0.1386_{-0.016-0.020}^{+0.0088+0.025}$ &
$0.1372_{-0.014-0.019}^{+0.0089+0.022}$ \\

$100\theta_\mathrm{MC}$ &
$1.04186_{-0.00025-0.00047}^{+0.00023+0.00047}$ &
$1.04190_{-0.00023-0.00045}^{+0.00023+0.00046}$ &
$1.04192_{-0.00024-0.00045}^{+0.00023+0.00048}$ &
$1.04192_{-0.00023-0.00045}^{+0.00023+0.00045}$ \\

$\tau$ &
$0.0572_{-0.0074-0.013}^{+0.0067+0.015}$ &
$0.0597_{-0.0079-0.015}^{+0.0070+0.015}$ &
$0.0591_{-0.0075-0.014}^{+0.0067+0.015}$ &
$0.0598_{-0.0074-0.014}^{+0.0068+0.015}$ \\

$n_\mathrm{s}$ &
$0.9666_{-0.0036-0.0069}^{+0.0036+0.0068}$ &
$0.9680_{-0.0034-0.0068}^{+0.0035+0.0066}$ &
$0.9678_{-0.0035-0.0070}^{+0.0035+0.0069}$ &
$0.9681_{-0.0037-0.0067}^{+0.0036+0.0070}$ \\

$\ln(10^{10} A_\mathrm{s})$ &
$3.044_{-0.014-0.026}^{+0.014+0.027}$ &
$3.049_{-0.014-0.028}^{+0.014+0.030}$ &
$3.047_{-0.015-0.028}^{+0.014+0.030}$ &
$3.048_{-0.015-0.027}^{+0.014+0.028}$ \\

$w_0$ &
$-1.103_{-0.044-0.098}^{+0.066+0.10}$ &
$>-1.078>-1.143$ &
$>-1.081>-1.148$ &
$>-1.067>-1.129$ \\

$\xi$ &
$>-0.0805>-0.1441$ &
$-0.069_{-0.019}^{+0.069}>-0.1468$ &
$-0.071_{-0.021}^{+0.067}>-0.1609$ &
$-0.066_{-0.024}^{+0.056}>-0.1417$ \\

$\Omega_\mathrm{m}$ &
$0.334_{-0.035-0.045}^{+0.021+0.054}$ &
$0.349_{-0.034-0.048}^{+0.025+0.053}$ &
$0.350_{-0.037-0.049}^{+0.022+0.058}$ &
$0.349_{-0.030-0.046}^{+0.023+0.050}$ \\

$\sigma_8$ &
$0.723_{-0.049-0.10}^{+0.070+0.094}$ &
$0.706_{-0.055-0.096}^{+0.060+0.094}$ &
$0.704_{-0.052-0.10}^{+0.065+0.095}$ &
$0.708_{-0.054-0.093}^{+0.054+0.090}$ \\

$H_0$ [Km/s/Mpc] &
$69.18_{-0.94-1.7}^{+0.87+1.8}$ &
$67.93_{-0.54-1.1}^{+0.52+1.1}$ &
$67.93_{-0.63-1.3}^{+0.63+1.2}$ &
$67.80_{-0.49-1.1}^{+0.54+1.0}$ \\

$S_8$ &
$0.760_{-0.025-0.057}^{+0.037+0.050}$ &
$0.759_{-0.028-0.055}^{+0.033+0.051}$ &
$0.758_{-0.027-0.055}^{+0.035+0.050}$ &
$0.761_{-0.025-0.054}^{+0.031+0.048}$ \\

$r_{\rm{drag}}$ [Mpc] &
$147.68_{-0.21-0.42}^{+0.21+0.41}$ &
$147.78_{-0.20-0.40}^{+0.20+0.40}$ &
$147.77_{-0.19-0.39}^{+0.20+0.38}$ &
$147.79_{-0.20-0.40}^{+0.20+0.42}$ \\

\hline
$\xi~(\sigma)$ & $1.4$ & $1.6$ & $1.6$ & $1.7$ \\
$\Delta\chi^2_{\rm min}$ & $-0.8$ & $-0.3$ & $-0.4$ & $-0.2$ \\
$\rm{ln}\mathcal{B}_{ij}$ & $-1.4$ & $-2.9$ & $-3.1$ & $-3.2$ \\
\hline

$\Omega_\mathrm{b} h^2$ &
$0.02225_{-0.00012-0.00025}^{+0.00012+0.00026}$ &
$0.02228_{-0.00013-0.00024}^{+0.00012+0.00026}$ &
$0.02227_{-0.00012-0.00025}^{+0.00012+0.00025}$ &
$0.02227_{-0.00012-0.00023}^{+0.00012+0.00024}$ \\

$\Omega_\mathrm{c} h^2$ &
$0.147_{-0.011-0.024}^{+0.016+0.021}$ &
$0.1518_{-0.0087-0.019}^{+0.011+0.018}$ &
$0.1529_{-0.0081-0.020}^{+0.012+0.017}$ &
$0.1510_{-0.0093-0.021}^{+0.013+0.019}$ \\

$100\theta_\mathrm{MC}$ &
$1.04176_{-0.00024-0.00047}^{+0.00024+0.00046}$ &
$1.04176_{-0.00022-0.00047}^{+0.00022+0.00045}$ &
$1.04175_{-0.00024-0.00046}^{+0.00024+0.00047}$ &
$1.04174_{-0.00022-0.00044}^{+0.00022+0.00045}$ \\

$\tau$ &
$0.0496_{-0.0070-0.014}^{+0.0070+0.014}$ &
$0.0487_{-0.0071-0.016}^{+0.0073+0.015}$ &
$0.0489_{-0.0074-0.015}^{+0.0076+0.015}$ &
$0.0488_{-0.0071-0.016}^{+0.0073+0.015}$ \\

$n_\mathrm{s}$ &
$0.9649_{-0.0035-0.0072}^{+0.0036+0.0066}$ &
$0.9655_{-0.0035-0.0068}^{+0.0036+0.0068}$ &
$0.9656_{-0.0036-0.0066}^{+0.0034+0.0069}$ &
$0.9652_{-0.0036-0.0072}^{+0.0036+0.0072}$ \\

$\ln(10^{10} A_\mathrm{s})$ &
$3.028_{-0.014-0.028}^{+0.014+0.027}$ &
$3.026_{-0.015-0.030}^{+0.014+0.029}$ &
$3.026_{-0.014-0.031}^{+0.015+0.029}$ &
$3.026_{-0.015-0.031}^{+0.015+0.030}$ \\

$\sum m_\mathrm{\nu,eff}$ &
$-0.142_{-0.082-0.15}^{+0.074+0.17}$ &
$-0.180_{-0.070-0.17}^{+0.079+0.15}$ &
$-0.184_{-0.068-0.15}^{+0.076+0.14}$ &
$-0.178_{-0.068-0.16}^{+0.084+0.17}$ \\

$w_0$ &
$>-1.074>-1.133$ &
$>-1.058>-1.100$ &
$>-1.051>-1.100$ &
$>-1.053>-1.097$ \\

$\xi$ &
$-0.094_{-0.053-0.074}^{+0.046+0.089}$ &
$-0.112_{-0.040-0.067}^{+0.040+0.071}$ &
$-0.117_{-0.044-0.068}^{+0.037+0.075}$ &
$-0.110_{-0.044-0.075}^{+0.040+0.081}$ \\

$\Omega_\mathrm{m}$ &
$0.356_{-0.023-0.056}^{+0.036+0.048}$ &
$0.371_{-0.017-0.043}^{+0.025+0.037}$ &
$0.376_{-0.018-0.046}^{+0.028+0.039}$ &
$0.370_{-0.018-0.047}^{+0.027+0.040}$ \\

$\sigma_8$ &
$0.701_{-0.069-0.085}^{+0.038+0.11}$ &
$0.680_{-0.046-0.069}^{+0.031+0.079}$ &
$0.674_{-0.049-0.069}^{+0.029+0.083}$ &
$0.682_{-0.051-0.073}^{+0.031+0.087}$ \\

$H_0$ [Km/s/Mpc] &
$68.18_{-0.94-1.8}^{+0.93+1.8}$ &
$67.54_{-0.63-1.2}^{+0.63+1.3}$ &
$67.30_{-0.73-1.3}^{+0.73+1.4}$ &
$67.47_{-0.61-1.3}^{+0.63+1.3}$ \\

$S_8$ &
$0.761_{-0.033-0.049}^{+0.026+0.055}$ &
$0.754_{-0.025-0.043}^{+0.022+0.045}$ &
$0.752_{-0.026-0.043}^{+0.021+0.046}$ &
$0.755_{-0.027-0.044}^{+0.022+0.047}$ \\

$r_{\rm{drag}}$ [Mpc] &
$147.61_{-0.21-0.42}^{+0.21+0.40}$ &
$147.62_{-0.21-0.42}^{+0.21+0.41}$ &
$147.66_{-0.22-0.42}^{+0.22+0.44}$ &
$147.64_{-0.21-0.41}^{+0.21+0.42}$ \\

\hline
$\xi~(\sigma)$ & $2.1$ & $3.0$ & $3.0$ & $2.7$ \\
$\Delta\chi^2_{\rm min}$ & $-0.4$ & $-4.4$ & $-5.3$ & $-6.2$ \\
$\rm{ln}\mathcal{B}_{ij}$ & $-0.9$ & $-2.0$ & $-1.8$ & $-1.3$ \\
\hline
\end{tabular}}

\caption{68\% and 95\% CL constraints on the \textbf{IDEphan}
scenario are presented for the CMB+DESI,
CMB+DESI+PantheonPlus, CMB+DESI+Union3, and
CMB+DESI+DES-Dovekie datasets. The upper half of the table
corresponds to the scenario with the neutrino mass fixed,
while the lower half corresponds to the scenario including
the effective neutrino mass as a free parameter.
We also report the statistical significance of a nonzero
coupling ($\xi$), obtained from its one-dimensional Gaussian
posterior using Wilks' theorem~\cite{Wilks:1938dza}.
This quantifies the statistical evidence for a nonzero value
of the coupling parameter, i.e., the deviation of $\xi$ from zero. }
\label{tab:IDEphan}
\end{table*}
\end{center}
\endgroup
\begin{figure*}
	\includegraphics[width=0.8\textwidth]{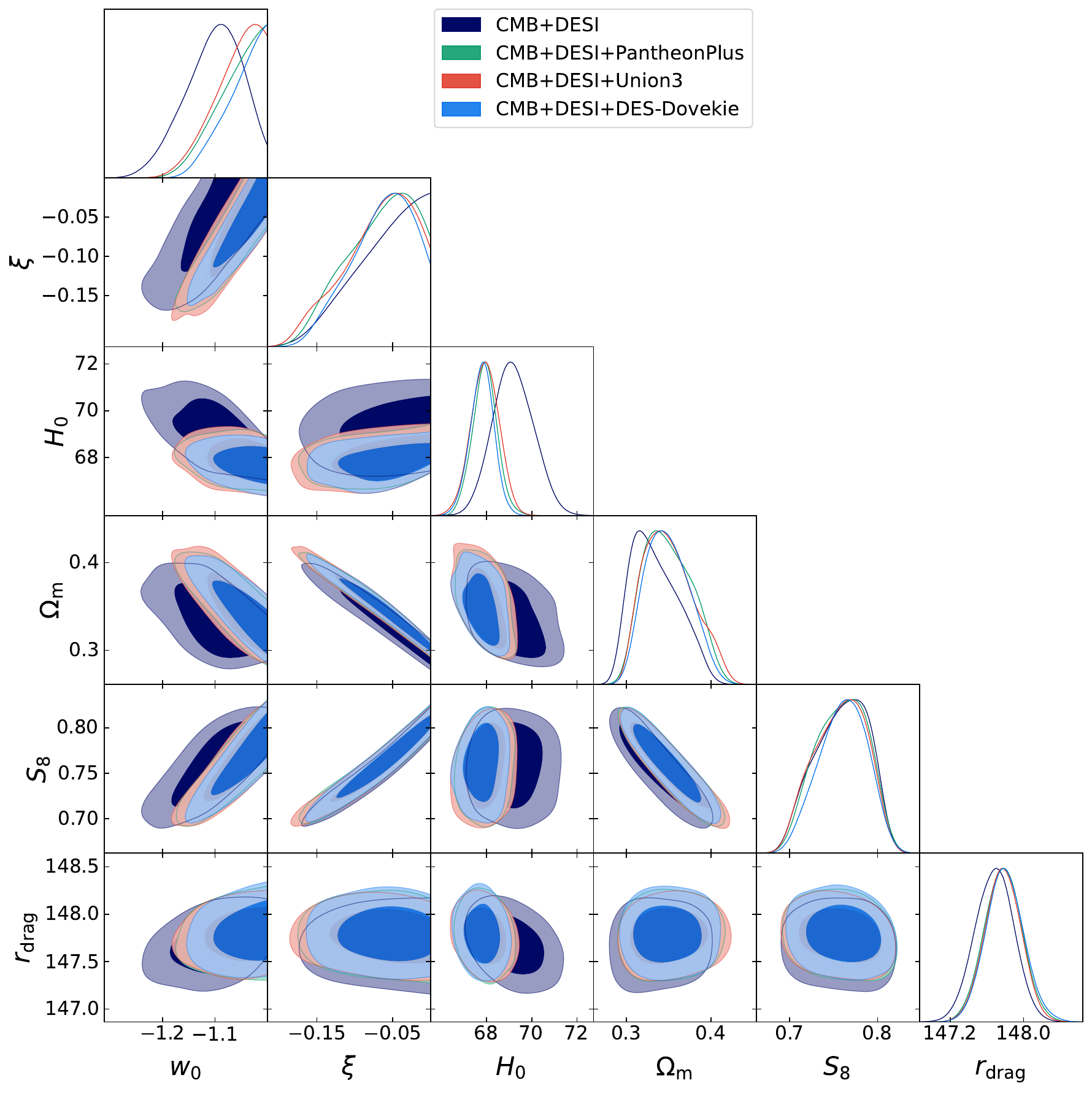}
	\caption{One-dimensional posterior distributions and
two-dimensional joint contours for the most relevant parameters
of the \textbf{IDEphan} scenario with fixed neutrino mass,
using different combinations of cosmological measurements.}
	\label{fig:IDEphan}
\end{figure*}
\begin{figure*}
	\includegraphics[width=0.8\textwidth]{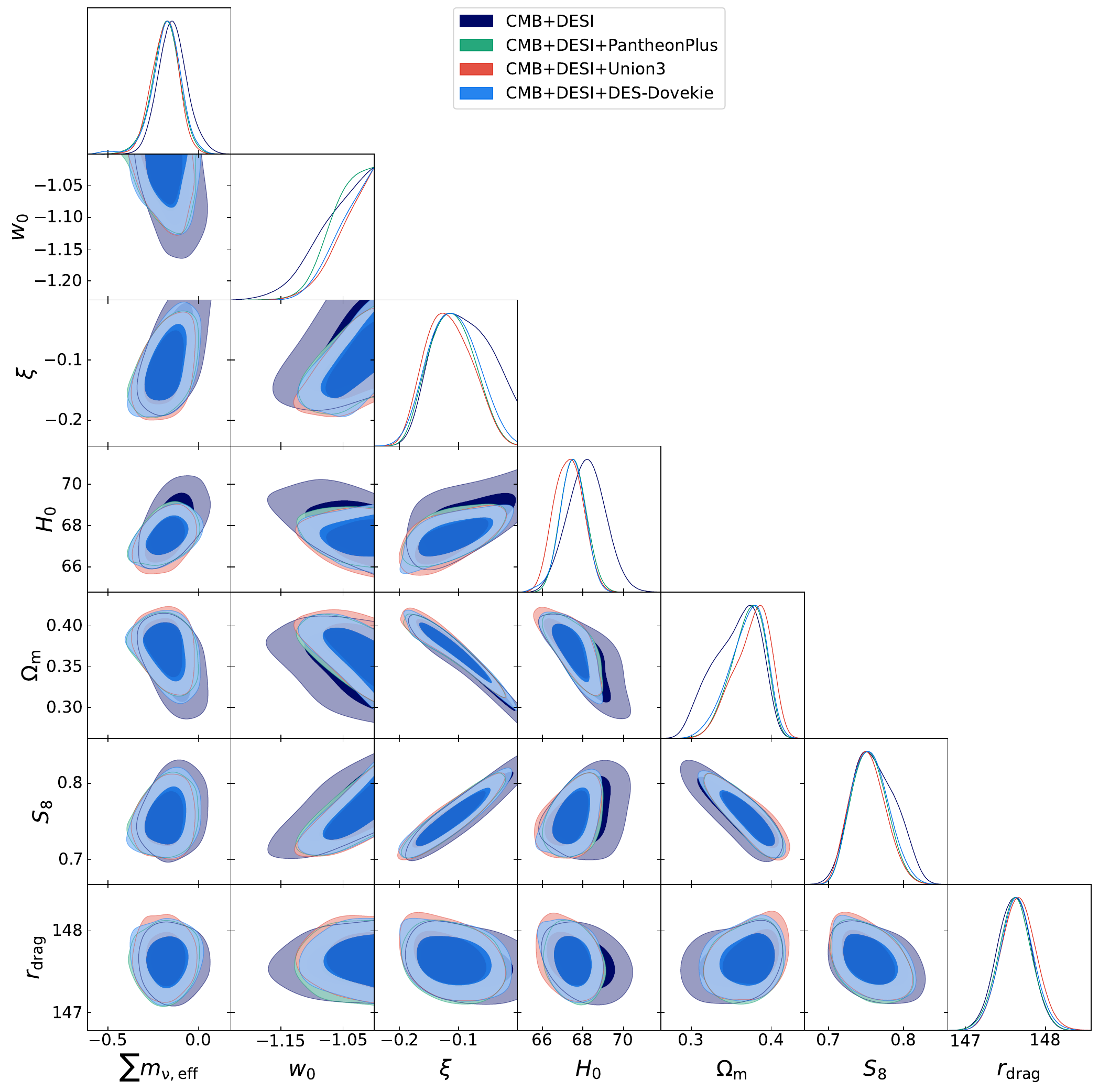}
	\caption{One-dimensional posterior distributions and
two-dimensional joint contours for the most relevant parameters
of the \textbf{IDEphan} scenario, using different combinations
of cosmological measurements.}
	\label{fig:IDEphan_NegMnu}
\end{figure*}
\begingroup
\begin{center}
	\begin{table*}
		\scalebox{0.85}{				
			\begin{tabular}{ccccccccc}
				\hline
				Parameters & CMB+DESI & CMB+DESI+PantheonPlus & CMB+DESI+Union3 & CMB+DESI+DES-Dovekie\\ \hline
				
				$\Omega_\mathrm{b} h^2$ & $0.02229_{-0.00013-0.00025}^{+0.00012+0.00026}$ & $0.02232_{-0.00012-0.00023}^{+0.00011+0.00024}$ & $0.02234_{-0.00012-0.00024}^{+0.00012+0.00024}$ & $0.02232_{-0.00013-0.00024}^{+0.00012+0.00025}$ \\
				
				$\Omega_\mathrm{c} h^2$ & $0.048_{-0.035-0.039}^{+0.017+0.049}$ & $0.055_{-0.039-0.044}^{+0.019+0.052}$ & $0.054_{-0.043-0.045}^{+0.018+0.058}$ & $0.056_{-0.044-0.045}^{+0.020+0.055}$ \\
				
				$100\theta_\mathrm{MC}$ & $1.04193_{-0.00024-0.00046}^{+0.00023+0.00048}$ & $1.04198_{-0.00024-0.00046}^{+0.00024+0.00044}$ & $1.04199_{-0.00024-0.00043}^{+0.00022+0.00046}$ & $1.04197_{-0.00023-0.00045}^{+0.00023+0.00044}$ \\
				
				$\tau$ & $0.0557_{-0.0076-0.014}^{+0.0066+0.015}$ & $0.0589_{-0.0076-0.014}^{+0.0070+0.015}$ & $0.0588_{-0.0076-0.014}^{+0.0065+0.016}$ & $0.0590_{-0.0074-0.014}^{+0.0072+0.015}$ \\
				
				$n_\mathrm{s}$ & $0.9674_{-0.0035-0.0074}^{+0.0037+0.0069}$ & $0.9690_{-0.0033-0.0064}^{+0.0033+0.0068}$ & $0.9689_{-0.0034-0.0070}^{+0.0034+0.0071}$ & $0.9690_{-0.0037-0.0070}^{+0.0036+0.0073}$ \\
				
				$\ln(10^{10} A_\mathrm{s})$ & $3.041_{-0.015-0.026}^{+0.013+0.029}$ & $3.047_{-0.015-0.028}^{+0.014+0.030}$ & $3.047_{-0.015-0.028}^{+0.013+0.030}$ & $3.047_{-0.015-0.028}^{+0.015+0.029}$ \\
				
				$w_0$ & $-0.870_{-0.057-0.13}^{+0.073+0.097}$ & $-0.835_{-0.051-0.14}^{+0.082+0.11}$ & $-0.834_{-0.050-0.15}^{+0.093+0.11}$ & $-0.832_{-0.050-0.14}^{+0.094+0.11}$ \\
				
				$\xi$ & $0.176_{-0.033-0.11}^{+0.073+0.083}$ & $0.157_{-0.038-0.13}^{+0.085+0.094}$ & $0.159_{-0.036}^{+0.090}<0.2428$ & $0.154_{-0.041-0.14}^{+0.093+0.097}$ \\
				
				$\Omega_\mathrm{m}$ & $0.147_{-0.072-0.082}^{+0.034+0.11}$ & $0.170_{-0.084-0.097}^{+0.044+0.11}$ & $0.167_{-0.091-0.098}^{+0.041+0.13}$ & $0.172_{-0.094-0.10}^{+0.046+0.12}$ \\
				
				$\sigma_8$ & $1.88_{-0.92-1.0}^{+0.42+1.4}$ & $1.66_{-0.83-0.89}^{+0.29+1.3}$ & $1.73_{-0.94-0.99}^{+0.34+1.4}$ & $1.69_{-0.90-0.94}^{+0.34+1.4}$ \\
				
				$H_0~[{\rm Km/s/Mpc}]$ & $69.32_{-0.96-1.7}^{+0.86+1.8}$ & $67.92_{-0.57-1.1}^{+0.56+1.2}$ & $67.95_{-0.65-1.3}^{+0.65+1.4}$ & $67.74_{-0.50-0.98}^{+0.51+1.0}$ \\
				
				$S_8$ & $1.21_{-0.30-0.36}^{+0.17+0.42}$ & $1.14_{-0.30-0.33}^{+0.14+0.43}$ & $1.16_{-0.34-0.37}^{+0.15+0.44}$ & $1.15_{-0.33-0.35}^{+0.31+0.44}$ \\
				
				$r_{\rm{drag}}~[{\rm Mpc}]$ & $147.70_{-0.22-0.41}^{+0.21+0.42}$ & $147.80_{-0.20-0.38}^{+0.19+0.37}$ & $147.79_{-0.20-0.40}^{+0.20+0.39}$ & $147.83_{-0.21-0.40}^{+0.20+0.40}$ \\
				
				\hline
				$\xi ~ (\sigma) $ & $3.1$ & $2.5$ & $2.3$ & $2.2$\\
				
				$\Delta\chi^2_{\rm min}$ & $-2.1$ & $-0.2$ & $-0.3$ & $-0.6$ \\
				
				$\rm{ln}\mathcal{B}_{ij}$ & $-0.6$ & $-1.7$ & $-1.3$ & $-1.6$ \\
								
				\hline 
								
				$\Omega_\mathrm{b} h^2$ & $0.02230_{-0.00011-0.00025}^{+0.00013+0.00022}$ & $0.02229_{-0.00011-0.00023}^{+0.00012+0.00022}$ & $0.02230_{-0.00011-0.00024}^{+0.00011+0.00022}$ & $0.02229_{-0.00011-0.00023}^{+0.00011+0.00024}$ \\
				
				$\Omega_\mathrm{c} h^2$ & $0.063_{-0.028-0.056}^{+0.031+0.048}$ & $0.067_{-0.026-0.048}^{+0.027+0.048}$ & $0.073_{-0.026-0.044}^{+0.026+0.045}$ & $0.071_{-0.027-0.047}^{+0.029+0.047}$ \\
				
				$100\theta_\mathrm{MC}$ & $1.04185_{-0.00024-0.00044}^{+0.00024+0.00045}$ & $1.04185_{-0.00023-0.00044}^{+0.00023+0.00046}$ & $1.04183_{-0.00023-0.00044}^{+0.00024+0.00045}$ & $1.04186_{-0.00023-0.00046}^{+0.00023+0.00046}$ \\
				
				$\tau$ & $0.0513_{-0.0075-0.016}^{+0.0077+0.016}$ & $0.0504_{-0.0075-0.016}^{+0.0076+0.015}$ & $0.0500_{-0.0070-0.016}^{+0.0079+0.015}$ & $0.0502_{-0.0070-0.016}^{+0.0077+0.014}$ \\
				
				$n_\mathrm{s}$ & $0.9669_{-0.0038-0.0073}^{+0.0037+0.0073}$ & $0.9667_{-0.0034-0.0074}^{+0.0035+0.0068}$ & $0.9671_{-0.0036-0.0071}^{+0.0036+0.0068}$ & $0.9667_{-0.0036-0.0072}^{+0.0037+0.0073}$ \\
				
				$\ln(10^{10} A_\mathrm{s})$ & $3.032_{-0.015-0.032}^{+0.015+0.031}$ & $3.030_{-0.014-0.033}^{+0.016+0.029}$ & $3.030_{-0.015-0.031}^{+0.015+0.029}$ & $3.030_{-0.014-0.030}^{+0.015+0.028}$ \\
				
				$\sum m_\mathrm{\nu,eff}~[{\rm eV}]$ & $-0.099_{-0.078-0.15}^{+0.081+0.15}$ & $-0.139_{-0.069-0.14}^{+0.067+0.15}$ & $-0.162_{-0.074-0.14}^{+0.067+0.15}$ & $-0.146_{-0.065-0.13}^{+0.062+0.14}$ \\
				
				$w_0$ & $-0.838_{-0.057-0.13}^{+0.070+0.11}$ & $-0.819_{-0.043-0.11}^{+0.066+0.092}$ & $-0.815_{-0.046-0.11}^{+0.069+0.098}$ & $-0.823_{-0.048-0.12}^{+0.073+0.098}$ \\
				
				$\xi$ & $0.138_{-0.068-0.12}^{+0.068+0.11}$ & $0.127_{-0.054-0.12}^{+0.065+0.095}$ & $0.112_{-0.055-0.11}^{+0.066+0.088}$ & $0.119_{-0.057-0.12}^{+0.070+0.091}$ \\
				
				$\Omega_\mathrm{m}$ & $0.180_{-0.066-0.11}^{+0.065+0.11}$ & $0.192_{-0.055-0.10}^{+0.059+0.10}$ & $0.207_{-0.058-0.095}^{+0.055+0.10}$ & $0.199_{-0.059-0.099}^{+0.060+0.10}$ \\
				
				$\sigma_8$ & $1.58_{-0.72-0.86}^{+0.15+1.6}$ & $1.43_{-0.55-0.68}^{+0.17+0.93}$ & $1.32_{-0.47-0.54}^{+0.13+0.77}$ & $1.37_{-0.52-0.60}^{+0.16+0.94}$ \\
				
				$H_0~[{\rm Km/s/Mpc}]$ & $68.35_{-0.98-1.9}^{+0.98+1.9}$ & $67.74_{-0.61-1.1}^{+0.57+1.2}$ & $67.37_{-0.71-1.3}^{+0.67+1.4}$ & $67.69_{-0.53-1.0}^{+0.54+1.1}$ \\
				
				$S_8$ & $1.115_{-0.26-0.32}^{+0.090+0.50}$ & $1.073_{-0.20-0.27}^{+0.088+0.33}$ & $1.034_{-0.18-0.22}^{+0.072+0.28}$ & $1.053_{-0.20-0.25}^{+0.085+0.33}$ \\
				
				$r_{\rm{drag}}$ Mpc & $147.60_{-0.22-0.39}^{+0.21+0.43}$ & $147.62_{-0.21-0.42}^{+0.22+0.41}$ & $147.64_{-0.20-0.40}^{+0.20+0.39}$ & $147.62_{-0.21-0.41}^{+0.22+0.42}$ \\
				
				\hline
				
				$\xi ~ (\sigma) $  & $2.2$ & $2.2$ & $2.0$ & $2.0$ \\
				
				$\Delta\chi^2_{\rm min}$ & $-0.1$ & $-4.1$ & $-5.2$ & $-6.1$ \\
				
				$\rm{ln}\mathcal{B}_{ij}$ & $-0.8$ & \boldmath{$+1.1$} & \boldmath{$+1.0$} & \boldmath{$+1.5$} \\
				
				\hline                                                   
				
		\end{tabular}}                                                                
		\caption{68\% and 95\% CL constraints on the \textbf{IDEquin}
scenario are presented for the CMB+DESI,
CMB+DESI+PantheonPlus, CMB+DESI+Union3, and
CMB+DESI+DES-Dovekie datasets. The upper half of the table
corresponds to the scenario with the neutrino mass fixed,
while the lower half corresponds to the scenario including
the effective neutrino mass as a free parameter.
We also report the statistical significance of a nonzero
coupling ($\xi$), obtained from its one-dimensional Gaussian
posterior using Wilks' theorem~\cite{Wilks:1938dza}.
This quantifies the statistical evidence for a nonzero value
of the coupling parameter, i.e., the deviation of $\xi$ from zero.}
		\label{tab:IDEquin}                   
	\end{table*}                                     
\end{center}
\endgroup
\begin{figure*}
	\includegraphics[width=0.8\textwidth]{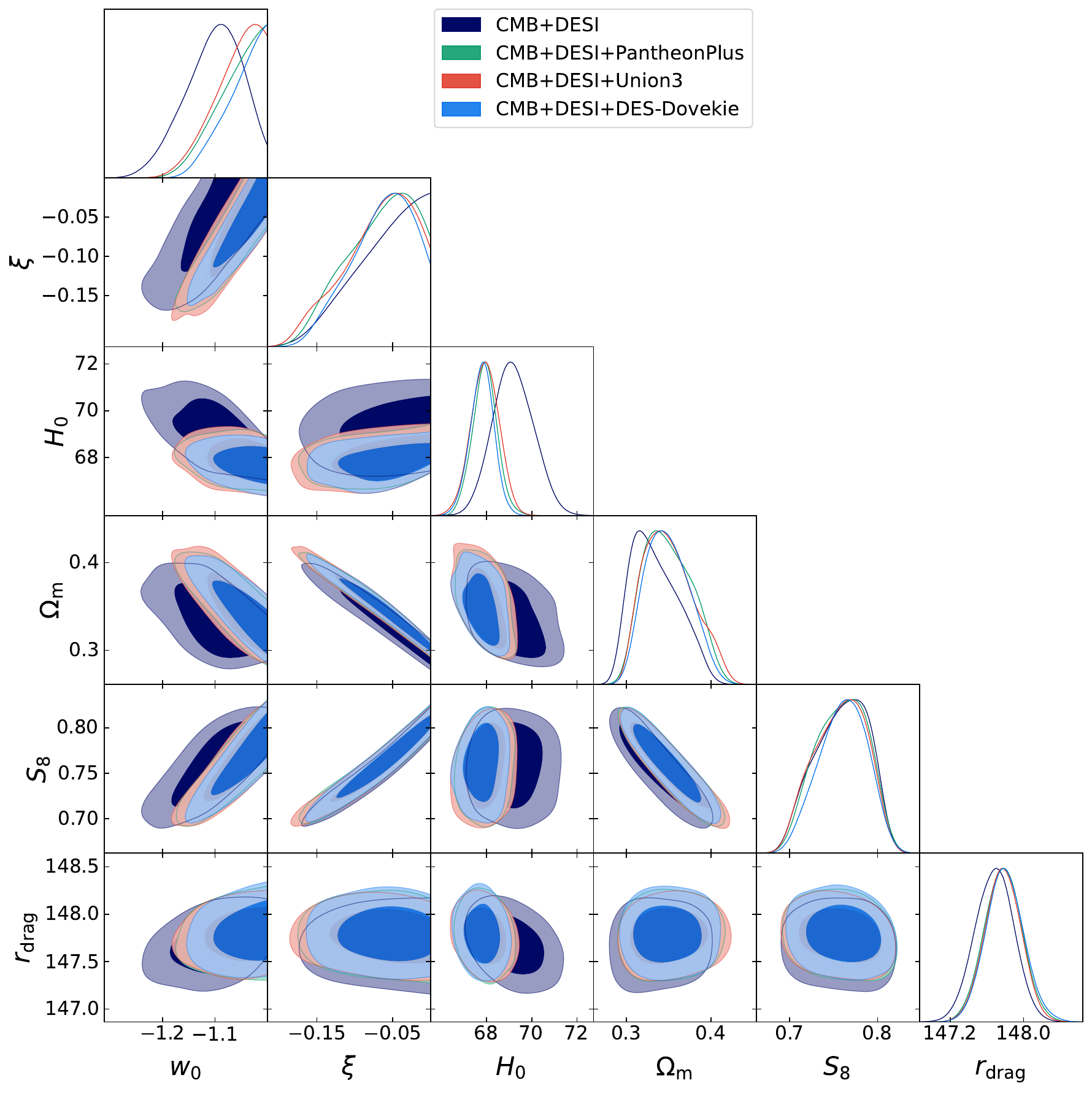}
	\caption{One-dimensional posterior distributions and
two-dimensional joint contours for the most relevant parameters
of the \textbf{IDEquin} scenario with fixed neutrino mass,
using different combinations of cosmological measurements.}
	\label{fig:IDEquin}
\end{figure*}
\begin{figure*}
	\includegraphics[width=0.8\textwidth]{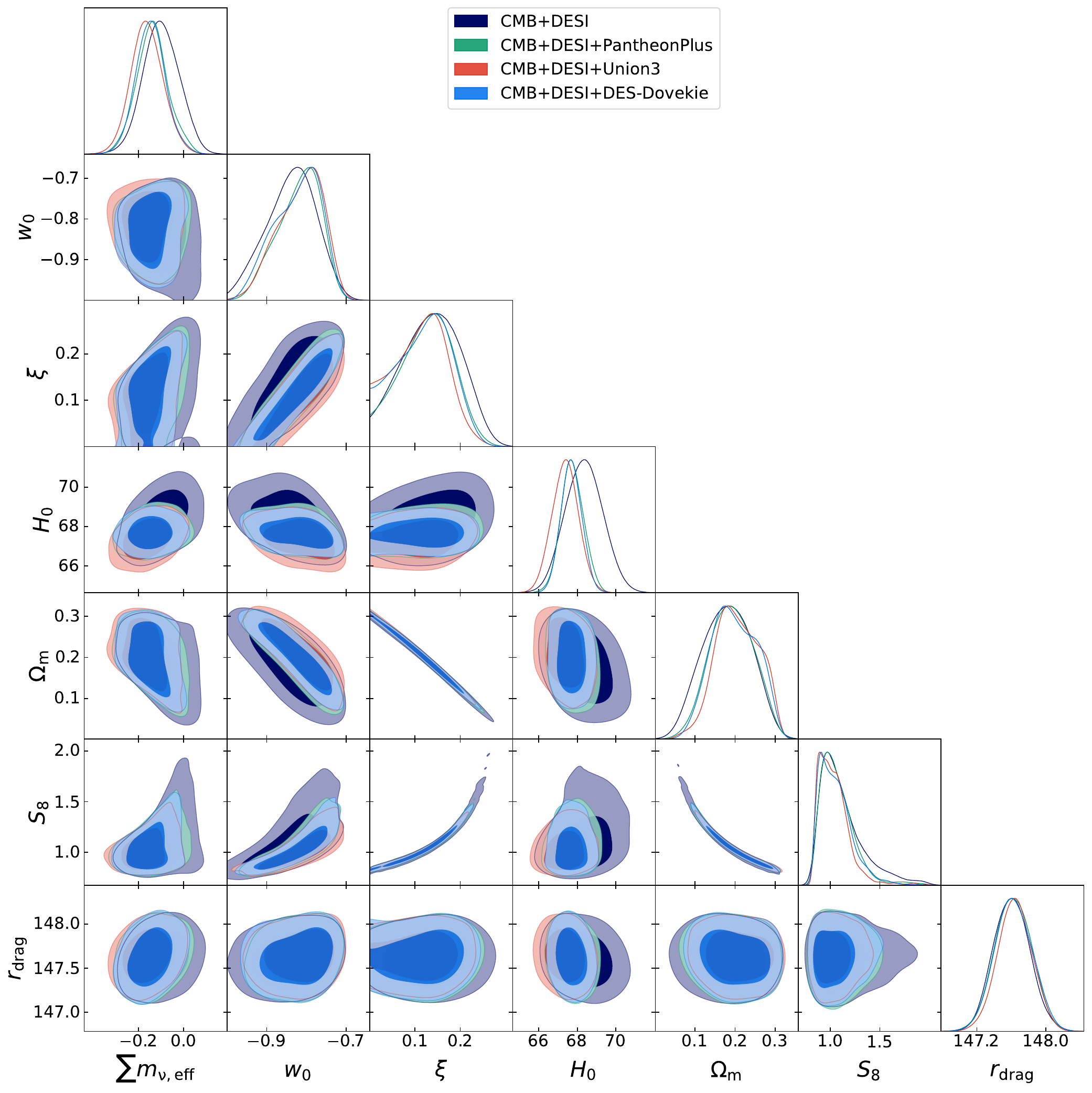}
	\caption{One-dimensional posterior distributions and
two-dimensional joint contours for the most relevant parameters
of the \textbf{IDEquin} scenario, using different combinations
of cosmological measurements.}
	\label{fig:IDEquin_NegMnu}
\end{figure*}

\section{Results}
\label{sec:4}

In this section, we describe the results for all three interacting
scenarios, namely, {\bf IVS}, {\bf IDEphan} and {\bf IDequin}. 
Considering four combined datasets, namely, CMB+DESI,
CMB+DESI+PantheonPlus, CMB+DESI+Union3, and
CMB+DESI+DES-Dovekie, we present the parameter constraints and
goodness of fit for all three interacting scenarios.  

\subsection{IVS}
The observational constraints on the {\bf IVS} and
{\bf IVS} + $\sum m_{\nu,{\rm eff}}$ models are summarized
in Table~\ref{tab:IVS}, while the corresponding
one-dimensional posterior distributions and two-dimensional joint
contours are presented in Figs.~\ref{fig:IVS}
and~\ref{fig:IVS_NegMnu}. The upper half of Table~\ref{tab:IVS}
reports the constraints obtained by fixing the total neutrino mass
to $\sum m_\nu=0.06$~eV, whereas the lower half presents the results
when the effective neutrino mass, $\sum m_{\nu,\rm eff}$, is allowed
to vary freely. In both analyses, we consider four different
combinations of cosmological datasets to investigate how the
inclusion of different low-redshift observations affects the
constraints on the model parameters and the statistical performance
of the model.

\subsubsection{Fixed total neutrino mass}

We first discuss the case in which the neutrino mass is fixed.
For the CMB+DESI dataset, the coupling parameter is constrained to
$\xi = 0.049^{+0.034}_{-0.037}$ ($68\%~{\rm CL}$), which shows
only a mild preference for a positive interaction. However, once
SN Ia observations are combined with CMB+DESI, the preferred value
of the coupling shifts towards zero and even slightly negative
values. For all the combined datasets, namely CMB+DESI+PantheonPlus,
CMB+DESI+Union3, and CMB+DESI+DES-Dovekie, $\xi=0$ is allowed
within the 68\% CL interval. In particular, the statistical
significance decreases from $1.4\sigma$ for CMB+DESI to
$0.50\sigma$, $0.33\sigma$, and $0.70\sigma$ after adding
PantheonPlus, Union3, and DES-Dovekie, respectively.

The remaining cosmological parameters exhibit the expected
behaviour. While the CMB+DESI combination allows relatively broad
ranges for parameters such as $\Omega_m$, $\sigma_8$, and $H_0$,
the addition of SN Ia observations tightens their posterior
distributions. In particular, SN Ia data favour a higher matter
density, shifting $\Omega_m$ from approximately $0.26$ to $0.31$.
Along the parameter degeneracies of this model, this shift is
accompanied by a decrease in $H_0$ from approximately $69.4$ to
$68~{\rm km\,s^{-1}\,Mpc^{-1}}$ and by a change in the preferred
sign of $\xi$. With our sign convention, positive $\xi$ corresponds
to energy transfer from dark matter to vacuum energy, whereas
negative $\xi$ corresponds to transfer from vacuum energy to dark
matter. Thus, the higher matter density preferred by SN Ia data
shifts the preferred direction of energy transfer towards
replenishing dark matter rather than depleting it. The clustering
amplitude also decreases from $\sigma_8\sim0.92$ for CMB+DESI
alone to $\sigma_8\sim0.79$ once SN Ia observations are included.
These shifts are driven by the additional constraining power of
the low-redshift distance measurements, while the coupling remains
consistent with zero.

Focusing on the model comparison statistics, we find that all
dataset combinations yield a small improvement in the minimum
$\chi^2_{\rm min}$ over $\Lambda$CDM with the same fixed neutrino
mass. For CMB+DESI, $\Delta\chi^2_{\rm min}=-1.9$, while adding
PantheonPlus, Union3, or DES-Dovekie yields
$\Delta\chi^2_{\rm min}=-0.3$, $-0.6$, and $-1.2$, respectively.
These small improvements alone do not establish statistical
support for the {\bf IVS} model.

The Bayesian evidence provides no support for the fixed-mass
{\bf IVS} model over $\Lambda$CDM. For CMB+DESI, the logarithmic
Bayes factor is $-0.5$, indicating an inconclusive comparison.
Adding PantheonPlus, Union3, or DES-Dovekie yields logarithmic
Bayes factors of $-1.6$, $-1.3$, and $-1.2$, respectively,
indicating weak support for $\Lambda$CDM after accounting for
the additional freedom introduced by the interaction parameter.

\subsubsection{Effective neutrino mass: free to vary}

We next consider the more general analysis in which the effective
neutrino mass is allowed to vary freely. As expected, introducing
$\sum m_{\nu,\rm eff}$ enlarges the cosmological parameter space
and introduces additional degeneracies, particularly involving
$\Omega_m$, $\xi$, and the effective neutrino mass. Nevertheless,
the recovered posterior distributions clearly show that the
preferred effective neutrino mass remains negative for all dataset
combinations, with central values around $-0.13$ to $-0.18$~eV.
Thus, {\bf IVS} does not remove the preference for negative neutrino
masses that appears once the positivity prior is relaxed.

Among the four dataset combinations, only CMB+DESI allows a positive
effective neutrino mass within the 95\% confidence interval.
However, its upper limit is approximately $0.03$~eV, still below
the minimum mass required by neutrino oscillation experiments.
This is also the dataset combination for which the interaction
itself receives essentially no statistical support
($\xi=-0.030^{+0.050}_{-0.044}$ at 68\% CL). Once any of
the SN Ia datasets is included, the posterior shifts further into
the negative-mass region, and the entire 95\% confidence interval
lies below zero. Therefore, none of these combinations reconciles
the cosmological constraints with the oscillation lower bounds
at this confidence level.

Allowing the effective neutrino mass to vary also modifies the
inferred behaviour of the interaction parameter. Along the parameter
degeneracies, negative effective neutrino masses allow the
preferred matter density to increase further. For the combinations
including SN Ia observations, $\Omega_m$ shifts from approximately
$0.31$ in the fixed-mass analysis to approximately $0.34$--$0.36$.
This higher matter density pushes $\xi$ towards more negative
values, favouring energy transfer from vacuum energy to dark
matter. The statistical significance increases compared to the
fixed-mass analysis, reaching approximately $1.9\sigma$,
$2.0\sigma$, and $2.5\sigma$ for the PantheonPlus, Union3, and
DES-Dovekie combinations, respectively. Thus, the enhanced
preference for a nonzero interaction is accompanied by a shift
towards negative effective neutrino masses and cannot be
interpreted as robust evidence for a physically viable
interacting vacuum scenario.

Finally, the model comparison diagnostics show a reduction
in $\chi^2_{\rm min}$ relative to $\Lambda$CDM with a free
effective neutrino mass, with $\Delta\chi^2_{\rm min}=-0.2$
for CMB+DESI and $-5.4$, $-5.9$, and $-5.4$ after adding
PantheonPlus, Union3, and DES-Dovekie, respectively.
The Bayesian comparison is inconclusive for CMB+DESI, with
$\ln\mathcal{B}_{ij}=0.1$, while the SN Ia combinations weakly
favour {\bf IVS}, with $\ln\mathcal{B}_{ij}=1.6$, $2.4$,
and $2.2$, respectively. Nevertheless, this statistical support
does not resolve the neutrino-mass tension, as the preferred
effective masses remain negative and the 95\% intervals remain
incompatible with oscillation bounds.

Overall, {\bf IVS} fails to provide a viable solution to the
neutrino-mass tension. When the neutrino mass is fixed, the
interaction parameter remains consistent with zero, the reduction
in $\chi^2_{\rm min}$ is small, and the Bayesian comparison is
inconclusive for CMB+DESI or weakly favours $\Lambda$CDM when
SN Ia data are included. Allowing the effective neutrino mass
to vary introduces a modest preference for a nonzero interaction
and weak Bayesian support for {\bf IVS} over free-mass
$\Lambda$CDM for the combinations including SN Ia data.
However, the preferred effective neutrino masses remain negative,
and none of the 95\% intervals accommodates the minimum mass
required by oscillation experiments. Therefore, the current
{\bf IVS} framework cannot simultaneously accommodate a
statistically preferred interaction and a neutrino mass
consistent with oscillation bounds.

\subsection{IDEphan}

The observational constraints on the {\bf IDEphan} and
{\bf IDEphan} + $\sum m_{\nu,\rm eff}$ models are summarized in
Table~\ref{tab:IDEphan}, while the corresponding one-dimensional
posterior distributions and two-dimensional joint contours are
presented in Figs.~\ref{fig:IDEphan}
and~\ref{fig:IDEphan_NegMnu}. The upper half of
Table~\ref{tab:IDEphan} reports the constraints obtained by fixing
the total neutrino mass to $\sum m_\nu=0.06$~eV, whereas the lower
half presents the results when the effective neutrino mass is
allowed to vary freely. As in the {\bf IVS} analysis, we consider four
combinations of cosmological datasets to investigate the effects
of including different SN Ia compilations.

\subsubsection{Fixed total neutrino mass}

We first discuss the case in which the neutrino mass is fixed.
For CMB+DESI, the dark energy equation of state is constrained to
$w_0=-1.103^{+0.066}_{-0.044}$ at 68\% CL, while the coupling
satisfies $\xi>-0.1441$ at 95\% CL. Adding SN Ia observations
shifts the equation of state towards the cosmological-constant
boundary, yielding only lower limits on $w_0$. The preferred
coupling remains negative, with central values around
$\xi\sim-0.07$ for the combinations including SN Ia data,
but the statistical significance of a departure from zero remains
modest, at approximately $1.4$--$1.7\sigma$ across the four
combinations. Thus, allowing a phantom equation of state does
not produce compelling evidence for a nonzero interaction
when the neutrino mass is fixed.

The remaining cosmological parameters show a coherent response
to the inclusion of SN Ia observations. The preferred matter
density increases from $\Omega_m\sim0.33$ for CMB+DESI to
$\Omega_m\sim0.35$, while $H_0$ decreases from approximately
$69.2$ to $68~{\rm km\,s^{-1}\,Mpc^{-1}}$. The clustering
amplitude decreases very mildly, from $\sigma_8\sim0.72$ to
$\sigma_8\sim0.71$, whereas $S_8$ remains close to $0.76$ for
all combinations. With our sign convention, the negative coupling
corresponds to energy transfer from dark energy to dark matter,
consistent with the higher late-time matter density preferred by
the combined datasets. However, the addition of SN Ia observations
does not substantially increase the statistical support for this
interaction.

The model comparison statistics likewise provide no substantial
support for the fixed-mass {\bf IDEphan} scenario. The differences
in $\chi^2_{\rm min}$ indicate only small improvements over
$\Lambda$CDM with the same fixed neutrino mass, with
$\Delta\chi^2_{\rm min}=-0.8$ for CMB+DESI and $-0.3$,
$-0.4$, and $-0.2$ after adding PantheonPlus, Union3,
and DES-Dovekie, respectively. The corresponding logarithmic
Bayes factors are $-1.4$, $-2.9$, $-3.1$, and $-3.2$.
Thus, the Bayesian evidence weakly favours $\Lambda$CDM
for CMB+DESI and moderately favours it for the combinations
including SN Ia observations.

\subsubsection{Effective neutrino mass: free to vary}

We next consider the analysis in which the effective neutrino
mass is allowed to vary freely. As in {\bf IVS}, the preferred
effective neutrino mass remains negative for every dataset
combination, with central values around $-0.14$~eV for CMB+DESI
and $-0.18$~eV after including SN Ia observations. Only CMB+DESI
allows positive effective masses within the 95\% confidence
interval, with an upper limit of approximately $0.03$~eV.
This is nevertheless below the minimum mass required by neutrino
oscillation experiments. For all combinations including SN Ia
observations, the entire 95\% interval lies below zero.
Thus, introducing a phantom equation of state alongside the
interaction does not reconcile the cosmological constraints with
the oscillation lower bounds.

Allowing negative effective neutrino masses also shifts the
preferred matter density higher, from $\Omega_m\sim0.33$ to
$\Omega_m\sim0.36$ for CMB+DESI and from $\Omega_m\sim0.35$
to $\Omega_m\sim0.37$--$0.38$ for the combinations including
SN Ia observations. Along the parameter degeneracies, this shift
pushes the coupling towards more negative values, with
$\xi\sim-0.09$ for CMB+DESI and $\xi\sim-0.11$ to $-0.12$
after including SN Ia data. The preferred energy transfer therefore
increasingly replenishes dark matter. Meanwhile, the dark energy
equation of state remains close to the cosmological-constant
boundary, with only lower limits on $w_0$, and $H_0$ decreases
to approximately $68.2~{\rm km\,s^{-1}\,Mpc^{-1}}$ for CMB+DESI
and $67.3$--$67.5~{\rm km\,s^{-1}\,Mpc^{-1}}$ for the
combinations including SN Ia observations.

The quoted statistical significance of a nonzero coupling
consequently rises to approximately $2.1\sigma$ for CMB+DESI and
$3.0\sigma$, $3.0\sigma$, and $2.7\sigma$ after adding
PantheonPlus, Union3, and DES-Dovekie, respectively. This enhanced
preference is accompanied by negative effective neutrino masses
and does not establish a physically viable solution to the
neutrino-mass tension. The clustering amplitude decreases further
to $\sigma_8\sim0.70$ for CMB+DESI and
$\sigma_8\sim0.67$--$0.68$ for the combinations including
SN Ia observations, while $S_8$ remains approximately
$0.75$--$0.76$.

Finally, the free-mass {\bf IDEphan} model yields a lower
$\chi^2_{\rm min}$ than free-mass $\Lambda$CDM, with
$\Delta\chi^2_{\rm min}=-0.4$ for CMB+DESI and $-4.4$,
$-5.3$, and $-6.2$ after adding PantheonPlus, Union3,
and DES-Dovekie, respectively. The Bayesian comparison is
inconclusive for CMB+DESI, with $\ln\mathcal{B}_{ij}=-0.9$.
The combinations including SN Ia observations yield
$\ln\mathcal{B}_{ij}=-2.0$, $-1.8$, and $-1.3$, respectively,
indicating weak support for free-mass $\Lambda$CDM.
Thus, the improvement in fit does not translate into Bayesian
support for {\bf IDEphan}.

Overall, {\bf IDEphan} fails to provide a viable solution to the
neutrino-mass tension. When the neutrino mass is fixed, the
interaction receives only modest statistical support and the
Bayesian evidence favours $\Lambda$CDM. Allowing the effective
neutrino mass to vary increases the quoted coupling significance
to approximately $3\sigma$, but the preferred masses remain
negative and the 95\% intervals remain incompatible with the
oscillation lower bounds. Consequently, the additional freedom
provided by the phantom equation of state does not resolve the
limitations found in the {\bf IVS} analysis.

\subsection{IDEquin}

The observational constraints on the \textbf{IDEquin} and
\textbf{IDEquin} + $\sum m_{\nu,\rm eff}$ models are summarized in
Table~\ref{tab:IDEquin}, while the corresponding one-dimensional
posterior distributions and two-dimensional joint contours are
presented in Figs.~\ref{fig:IDEquin}
and~\ref{fig:IDEquin_NegMnu}. The upper half of
Table~\ref{tab:IDEquin} reports the constraints obtained by fixing
the total neutrino mass to $\sum m_\nu=0.06$~eV, whereas the lower
half presents the results when the effective neutrino mass is
allowed to vary freely. As in the previous analyses, we consider
four combinations of cosmological datasets to investigate the
effects of including different SN Ia compilations.

\subsubsection{Fixed total neutrino mass}

We first discuss the case in which the neutrino mass is fixed.
For CMB+DESI, the dark energy equation of state is constrained to
$w_0=-0.870^{+0.073}_{-0.057}$ at 68\% CL, while the coupling is
$\xi=0.176^{+0.073}_{-0.033}$ at 68\% CL. The quoted statistical
significance of a nonzero coupling reaches approximately
$3.1\sigma$, the largest value obtained in the fixed-mass analyses
considered here. Adding SN Ia observations shifts the preferred
equation of state towards $w_0\sim-0.83$ and reduces the preferred
coupling to $\xi\sim0.15$--$0.16$. The corresponding significance
decreases to approximately $2.5\sigma$, $2.3\sigma$, and
$2.2\sigma$ after adding PantheonPlus, Union3, and DES-Dovekie,
respectively. Thus, SN Ia observations weaken the preference
for an interaction in the fixed-mass \textbf{IDEquin} scenario.

The remaining cosmological parameters respond consistently to
these shifts. The preferred matter density increases from
$\Omega_m\sim0.15$ for CMB+DESI to $\Omega_m\sim0.17$
after including SN Ia observations, while $H_0$ decreases from
approximately $69.3$ to $68~{\rm km\,s^{-1}\,Mpc^{-1}}$.
With our sign convention, positive $\xi$ corresponds to energy
transfer from dark matter to dark energy. The higher matter
density preferred by SN Ia data therefore shifts the coupling
towards smaller positive values, reducing the preferred depletion
of dark matter without reversing the direction of energy transfer.
The inferred clustering amplitude is large and poorly constrained,
with $\sigma_8\sim1.9$ for CMB+DESI and $\sigma_8\sim1.7$
after including SN Ia observations. The corresponding $S_8$
values remain around $1.1$--$1.2$, with broad asymmetric
uncertainties.

The fixed-mass model comparison yields negative
$\Delta\chi^2_{\rm min}$ values for every dataset combination.
However, the improvement is modest:
$\Delta\chi^2_{\rm min}=-2.1$ for CMB+DESI and less than unity
in magnitude after including SN Ia observations. The logarithmic
Bayes factor is $-0.6$ for CMB+DESI, indicating an inconclusive
comparison. Adding PantheonPlus, Union3, or DES-Dovekie yields
logarithmic Bayes factors of $-1.7$, $-1.3$, and $-1.6$,
respectively, indicating weak support for $\Lambda$CDM.
Thus, the preference for a nonzero coupling within the fixed-mass
\textbf{IDEquin} parameter space does not translate into Bayesian
support for the model.

\subsubsection{Effective neutrino mass: free to vary}

We next allow the effective neutrino mass to vary freely.
The preferred masses remain negative for every dataset combination,
with central values around $-0.10$~eV for CMB+DESI and
$-0.14$ to $-0.16$~eV after including SN Ia observations.
The 95\% CL upper limits are approximately $0.052$~eV for
CMB+DESI and $0.011$~eV after adding PantheonPlus.
Both combinations permit positive effective masses, but their
upper limits remain below the minimum mass required by neutrino
oscillation experiments. For Union3 and DES-Dovekie, the upper
limits are approximately $-0.015$~eV and $-0.008$~eV,
respectively, so the entire 95\% interval lies below zero.
Consequently, none of these combinations reconciles the
cosmological constraints with the oscillation lower bounds
at this confidence level.

Allowing negative effective neutrino masses shifts the preferred
matter density higher, to $\Omega_m\sim0.18$ for CMB+DESI
and $\Omega_m\sim0.19$--$0.21$ for the combinations including
SN Ia observations. Along the parameter degeneracies, this
increase is accompanied by a reduction in the preferred coupling
to $\xi\sim0.14$ for CMB+DESI and $\xi\sim0.11$--$0.13$
after including SN Ia data. The coupling remains positive, but
the preferred transfer of energy out of dark matter is weaker.
The equation of state shifts towards $w_0\sim-0.84$ for
CMB+DESI and $w_0\sim-0.82$ for the SN Ia combinations.
Meanwhile, $H_0$ decreases to approximately
$68.4~{\rm km\,s^{-1}\,Mpc^{-1}}$ for CMB+DESI and
$67.4$--$67.7~{\rm km\,s^{-1}\,Mpc^{-1}}$ after including
SN Ia observations. The clustering amplitude also decreases,
although its posterior remains broad and asymmetric.

Unlike {\bf IVS} and {\bf IDEphan}, allowing the effective neutrino
mass to vary does not strengthen the interaction preference.
The quoted coupling significance remains approximately
$2.0$--$2.2\sigma$ across all four combinations.
The $\chi^2_{\rm min}$ is lower than that of free-mass
$\Lambda$CDM, with $\Delta\chi^2_{\rm min}=-4.1$, $-5.2$,
and $-6.1$ for the combinations including PantheonPlus, Union3,
and DES-Dovekie, respectively, compared with
$\Delta\chi^2_{\rm min}=-0.1$ for CMB+DESI.
The Bayesian comparison is inconclusive for CMB+DESI, with
$\ln\mathcal{B}_{ij}=-0.8$. Adding PantheonPlus, Union3,
or DES-Dovekie yields $\ln\mathcal{B}_{ij}=1.1$, $1.0$,
and $1.5$, respectively, indicating weak support for
\textbf{IDEquin} over free-mass $\Lambda$CDM. However,
this support does not restore compatibility with the neutrino
oscillation lower bounds.

Overall, \textbf{IDEquin} yields a lower $\chi^2_{\rm min}$
than the corresponding $\Lambda$CDM model for all combinations
considered. When the effective neutrino mass is varied, the
combinations including SN Ia data also weakly favour
\textbf{IDEquin} in Bayesian evidence. Although CMB+DESI and
CMB+DESI+PantheonPlus permit small positive effective neutrino
masses, their 95\% upper limits remain below the oscillation
minimum. The other combinations exclude positive masses at
this confidence level. Therefore, despite its interaction
preference and additional freedom in the dark energy equation
of state, \textbf{IDEquin} fails to provide a viable solution
to the neutrino-mass tension.

\subsection{Summary}

We analyzed three dark matter-dark energy interaction models to
test their compatibility with the neutrino mass lower bounds
established by oscillation experiments. Our analysis complements
recent investigations of interacting vacuum models with a
non-negative neutrino mass prior, including a reported bound of
$\sum m_\nu<0.051$~eV at 95\% CL for an interaction proportional
to $H_0\rho_x$ and a flat prior
$\sum m_\nu/{\rm eV}\in[0,5]$~\cite{Li:2026ldf}.
For the interaction proportional to $H\rho_x$ considered here,
allowing the effective neutrino mass parameter to take negative
values reveals a negative preferred mass in all three scenarios
and for every dataset combination. None of the resulting 95\%
intervals reaches the minimum mass required by oscillation
experiments.

In {\bf IVS}, allowing the effective neutrino mass to vary increases
the preferred matter density and shifts the coupling towards
more negative values, favouring energy transfer from vacuum
energy to dark matter. The coupling significance reaches
approximately $2.5\sigma$ after including SN Ia observations,
but the preferred masses remain negative. Only CMB+DESI permits
positive effective masses at 95\% CL, with an upper limit of
approximately $0.03$~eV, still below the oscillation minimum.
The phantom extension exhibits the same qualitative behaviour:
the coupling significance increases to approximately $3\sigma$,
while positive masses are excluded at 95\% CL for every
combination including SN Ia observations. CMB+DESI again permits
only positive masses below approximately $0.03$~eV.

The quintessence regime differs in that the coupling remains
positive and becomes weaker when the effective neutrino mass
is allowed to vary. CMB+DESI and CMB+DESI+PantheonPlus permit
positive effective masses at 95\% CL, but their upper limits
of approximately $0.052$~eV and $0.011$~eV remain below the
oscillation minimum. The Union3 and DES-Dovekie combinations
exclude positive masses at this confidence level. For the
combinations including SN Ia observations, all three free-mass
interacting models yield lower $\chi^2_{\rm min}$ values than
free-mass $\Lambda$CDM. The Bayesian evidence weakly favours
{\bf IVS} and \textbf{IDEquin}, while weakly favouring
$\Lambda$CDM over {\bf IDEphan}. Nevertheless, none of these
scenarios resolves the neutrino-mass tension.

These extensions contain additional free parameters relative
to the corresponding $\Lambda$CDM baseline. Bayesian evidence
implements Occam's razor by penalizing the additional prior
volume unless the improvement in fit is sufficient to compensate
for it~\cite{Trotta:2008qt}. Consequently, a reduction in
$\chi^2_{\rm min}$ does not by itself establish support for the
extended model. For regular nested models under the null
hypothesis, an average reduction of approximately one unit in
$\chi^2_{\rm min}$ per additional free parameter is expected
from fitting statistical fluctuations alone~\cite{Wilks:1938dza}.
This expectation is not universal, particularly when the
reference model lies on a parameter boundary, as in the phantom
and quintessence branches considered here. For the adopted
priors, the free-mass {\bf IVS} and \textbf{IDEquin} models
overcome the Occam penalty sufficiently to receive weak Bayesian
support when SN Ia observations are included. However, this
support does not restore compatibility with the neutrino mass
lower bounds established by oscillation experiments.

These results demonstrate that the interaction and constant
dark energy equations of state considered here are insufficient
to resolve the neutrino-mass tension. They do not, however,
establish a general failure of coupled quintessence.
A recent analysis of a quintessence field with an exponential
potential and exponential dark matter coupling also finds a
negative preferred effective mass, but with broader uncertainties;
for its negative-coupling branch, the reported constraint is
$\sum m_{\nu,\rm eff}=-0.052^{+0.068}_{-0.10}$~eV
at 68\% CL~\cite{Wang:2026gvg}.
The contrast with our results shows that the neutrino mass
constraints depend on the dynamics of the dark energy field
and the form of its coupling, rather than on the presence
of a dark-sector interaction alone. Dynamical coupled quintessence~\cite{Bedroya:2025fwh,Li:2026xaz} 
therefore provides a more promising route to investigate than
the constant-equation-of-state scenarios analyzed here.

\section{Conclusions}
\label{sec:5}

The standard model of cosmology ($\Lambda$CDM) faces critical
evaluation in light of recent cosmological observations.
DESI data, when combined with CMB and SN Ia measurements,
reveal a preference for dynamical dark energy over a
cosmological constant, reaching a statistical significance
above $3\sigma$~\cite{DESI:2025zgx,DES:2025sig}.
These observations also expose a tension in the neutrino
sector~\cite{Elbers:2025vlz}. When the positive-mass prior is
relaxed within $\Lambda$CDM, cosmological analyses favour a
negative effective neutrino mass
parameter~\cite{Craig:2024tky,Green:2024xbb,Loverde:2024nfi,Elbers:2025vlz}.
This unphysical preference conflicts with the positive minimum
mass established by terrestrial neutrino oscillation
experiments~\cite{Esteban:2024eli}.

Allowing a time-varying dark energy equation of state changes
the expansion history and the associated parameter degeneracies,
widening the permitted neutrino mass range and restoring
compatibility with oscillation bounds~\cite{Elbers:2025vlz}.
Motivated by this interplay, we investigated whether
non-gravitational interactions between dark matter and dark
energy could achieve the same reconciliation. We considered
interacting vacuum energy and interacting dark energy with a
constant equation of state in the phantom and quintessence
regimes, allowing the effective neutrino mass parameter to
explore both positive and negative values.

Our analyses demonstrate that these scenarios fail to resolve
the neutrino-mass tension. None yields a 95\% interval that
accommodates the minimum mass required by oscillation experiments.
The presence of an interaction, even when accompanied by a
preference for nonzero coupling, a reduction in
$\chi^2_{\rm min}$, or weak Bayesian support over the corresponding
$\Lambda$CDM baseline, is therefore insufficient to recover a
physically acceptable neutrino sector. This distinguishes the
ability to fit cosmological observations from the ability to
reconcile them with independently established particle physics.

The failure of these scenarios places a concrete restriction
on phenomenological explanations of the DESI results:
introducing energy transfer within the dark sector does not
automatically reproduce the relaxation of neutrino mass bounds
obtained with dynamical dark energy. The relevant physics
depends on the evolution of the dark energy density and the
coupling, together with their effects on cosmological
perturbations. Consequently, our findings constrain the specific
interaction and constant-equation-of-state models examined
here, rather than excluding dark-sector interactions in general.

Dynamical coupled quintessence remains an important direction.
A model with an exponential potential and exponential coupling
to cold dark matter~\cite{Bedroya:2025fwh} yields broader effective neutrino mass
constraints than those obtained here, although its preferred
mass remains negative~\cite{Wang:2026gvg}.
Alternatively, a sign-switching cosmological
constant~\cite{Akarsu:2019hmw,Akarsu:2021fol,Akarsu:2022typ,Akarsu:2023mfb}
has been shown to recover a positive preferred effective neutrino
mass compatible with oscillation bounds by modifying the
expansion history over the relevant redshift
range~\cite{Kibris:2026cqq}. These results motivate further
investigation of the specific dark energy dynamics needed
to reconcile cosmological distances, structure formation,
and neutrino physics. Such investigations should be accompanied
by continued scrutiny of systematic uncertainties in the
combined datasets.

\section*{Acknowledgements}

W.Y. is supported by the National Natural Science Foundation of China under Grant Nos. 12547110 and 12175096. L.A.A. is supported by the U.S. National Science
Foundation (NSF Grant PHY-2412679), and he extends his appreciation to the
Harvard Swampland Initiative for their warm hospitality and for
providing a stimulating environment for productive discussions. E.D.V. is supported by a Royal Society Dorothy Hodgkin Research Fellowship. This article is based upon work from the COST Action CA21136 - ``Addressing observational tensions in cosmology with systematics and fundamental physics (CosmoVerse)'', supported by COST - ``European Cooperation in Science and Technology''.

\bibliographystyle{apsrev4-1}
\bibliography{biblio}


\end{document}